\documentclass[twocolumn]{aastex63}
\usepackage{natbib,twoopt,amsmath}
\usepackage{color}
\hypersetup{linkcolor=red,citecolor=green,filecolor=cyan,urlcolor=magenta}

\shorttitle{Circumstellar environment of MWC\,137}
\shortauthors{Kraus et al.}

\begin{document}
\title{Resolving the circumstellar environment of the Galactic B[e] supergiant star MWC\,137.\\
II. Nebular kinematics and stellar variability\footnote{Based on observations collected with 1) the Nordic Optical Telescope, operated by the Nordic Optical Telescope Scientific Association at the Observatorio del Roque de los Muchachos, La Palma, Spain, of the Instituto de Astrofisica de Canarias, 2) the Danish 1.54-m telescope at La Silla, Chile, and 3) the 6-m telescope of the Special Astrophysical Observatory (SAO), Russia.}}

\correspondingauthor{Michaela Kraus}
\email{michaela.kraus@asu.cas.cz}

\author[0000-0002-4502-6330]{Michaela Kraus}
\affil{Astronomical Institute, Czech Academy of Sciences, 
251\,65 Ond\v{r}ejov, Czech Republic}

\author[0000-0003-2196-9091]{Tiina Liimets}
\affil{Astronomical Institute, Czech Academy of Sciences, 
251\,65 Ond\v{r}ejov, Czech Republic}
\affiliation{Tartu Observatory, University of Tartu, Observatooriumi 1, 61602 T\~oravere,
Estonia}

\author[0000-0002-0507-9307]{Alexei Moiseev}
\affil{Special Astrophysical Observatory, Russian Academy of Sciences, Nizhnii Arkhyz 369167, Russia}

\author[0000-0002-6865-4321]{Julieta P. S\'{a}nchez Arias}
\affil{Astronomical Institute, Czech Academy of Sciences, 
251\,65 Ond\v{r}ejov, Czech Republic}

\author[0000-0001-5165-6331]{Dieter H. Nickeler}
\affil{Astronomical Institute, Czech Academy of Sciences, 
251\,65 Ond\v{r}ejov, Czech Republic}

\author{Lydia S. Cidale}
\affil{Departamento de Espectroscop\'ia, Facultad de Ciencias Astron\'omicas y Geof\'isicas, 
Universidad Nacional de La Plata,
Paseo del Bosque S/N, La Plata, B1900FWA, Buenos Aires, Argentina}
\affiliation{Instituto de Astrof\'isica de La Plata (CCT La Plata - CONICET, UNLP),
Paseo del Bosque S/N, La Plata, B1900FWA, Buenos Aires,  Argentina}

\author[0000-0003-3947-5946]{David Jones}
\affil{Instituto de Astrof\'{i}sica de Canarias, E-38205 La Laguna, Tenerife, Spain}
\affiliation{Departamento de Astrof\'{i}sica, Universidad de La Laguna, E-38206 La Laguna, Tenerife, Spain}

\begin{abstract}
The Galactic B[e] supergiant MWC\,137 is surrounded by a large-scale optical nebula. To shed light on 
the physical conditions and kinematics of the nebula, we analyze the optical forbidden emission lines 
\protect{[N\,\sc{ii}]}\,$\lambda\lambda$\,6548,6583 and 
\protect{[S\,\sc{ii}]}\,$\lambda\lambda$\,6716,6731 in long-slit spectra taken with ALFOSC at the 
Nordic Optical Telescope. The radial velocities display a complex behavior but, in general, the 
northern nebular features are predominantly approaching while the southern ones are mostly receding. 
The electron density shows strong variations across the nebula with values spreading from about zero 
to $\sim800$\,cm$^{-3}$. Higher densities are found closer to MWC~137 and in regions of intense 
emission, whereas in regions with high radial velocities the density decreases significantly. We also 
observe the entire nebula in the two \protect{[S\,\sc{ii}]} lines with the scanning Fabry-Perot 
interferometer attached to the 6-m telescope of the Special Astrophysical Observatory. These data 
reveal a new bow-shaped feature at PA $=225-245\degr$ and a distance 80$\arcsec$ from MWC~137. A new 
H$\alpha$ image has been taken with the Danish 1.54-m telescope on La Silla. No expansion or changes 
in the nebular morphology appear within 18.1 years. We derive a mass of $37^{+9}_{-5}\,M_{\sun}$ and 
an age of $4.7\pm0.8$\,Myr for MWC~137. Furthermore, we detect a period of 1.93\,d in the time series 
photometry collected with the TESS satellite, which could suggest stellar pulsations. Other, 
low-frequency variability is seen as well. Whether these signals are caused by internal gravity waves 
in the early-type star or by variability in the wind and circumstellar matter currently cannot be 
distinguished. 
\end{abstract}


\section{Introduction}

The class of B[e] supergiants (B[e]SG) consists of evolved, luminous massive stars, which display
indications of a circumstellar dusty disk and a bipolar line-driven wind \citep{1985A&A...143..421Z, 
1998A&A...340..117L}. Their main optical spectral characteristics are intense and broad Balmer line 
emission along with narrow emission lines of low-ionized and neutral metals of both permitted and 
forbidden transitions, in particular [O\,{\sc i}]. Moreover, many B[e]SGs display intense emission 
of the Ca\,{\sc ii} infrared triplet lines and of [Ca\,{\sc ii}] \citep{2012MNRAS.423..284A, 
2019Galax...7...83K}. These optical forbidden emission lines have been identified as valuable tracers 
for the circumstellar gas disks of B[e]SGs and have been used to determine the kinematics within 
their emission regions \citep{2003A&A...408..257Z, 2007A&A...463..627K, 2010A&A...517A..30K, 
2016A&A...593A.112K, 2018MNRAS.480..320M, 2018A&A...612A.113T}. 

The spectral energy distribution of B[e]SGs displays strong infrared (IR) excess emission due to hot 
and warm circumstellar dust \citep{1976ApJ...210..666A, 1986A&A...163..119Z, 2009AJ....138.1003B, 
2010AJ....140..416B}, and their IR colors have been proven suitable to distinguish B[e]SGs from other 
types of massive stars, such as Luminous Blue Variables and massive pre-main sequence objects 
\citep{2013A&A...558A..17O, 2019Galax...7...83K, 2020AJ....160..166C}. Furthermore, the IR spectra of 
B[e]SGs provide complementary information about their environments. Most valuable in this regard is the 
molecular emission of CO \citep{1988ApJ...334..639M, 1996ApJ...470..597M, 2012A&A...548A..72C, 
2012MNRAS.426L..56O, 2012A&A...543A..77W, 2014ApJ...780L..10K} and its isotope $^{13}$CO 
\citep{2010MNRAS.408L...6L, 2013A&A...558A..17O, 2020MNRAS.493.4308K} from the circumstellar disks. 
The detection of an enrichment in the isotope is clear evidence that the disks have been formed from 
stellar wind material that has been enriched with chemically processed matter mixed to the stellar 
surface during the evolution of the star \citep{2009A&A...494..253K}. 

Emission from SiO has also been detected in a sample of B[e]SGs \citep{2015ApJ...800L..20K}. The 
kinematics of the molecular gas, obtained from high-resolution spectra, provides further information 
on the global disk dynamics. Under the assumption of strict Keplerian rotation, the combined velocity 
information from the optical forbidden lines and the molecular emission suggests that B[e]SGs are 
surrounded by either multiple ring structures or disks with strongly varying density in radial 
direction \citep{2018MNRAS.480..320M}. Moreover, the rings around some objects seem to display 
inhomogeneities as is inferred from the temporal variability of the emission features 
\citep{2016A&A...593A.112K, 2018A&A...612A.113T}.

The Galactic object MWC~137 (V1308 Ori, $\alpha$ = 06:18:45.52, $\delta$ = $+$15:16:52.24, J2000) is 
a peculiar emission-line star, surrounded by the optical nebula Sh~2-266. Its evolutionary state was 
long unclear. Its resemblance with massive pre-main sequence stars led to the inclusion of the object 
into catalogs of Herbig Be stars \citep[e.g.,][]{1992ApJ...397..613H, 1992ApJ...398..254B, 
1994A&AS..104..315T}, whereas the large scale environment with an optical nebula 
\citep{2008A&A...477..193M}, embedded in a dusty and cold molecular gas component 
\citep{2017AJ....154..186K} which are reminiscent of the bipolar lobes and equatorial rings seen 
in the B-type supergiants Sher~25 and SBW~1 \citep{1997ApJ...475L..45B, 2007AJ....134..846S}, 
questioned the young nature. 

Emission from the isotope $^{13}$CO has been detected from MWC~137 in medium-resolution K-band 
spectra \citep{2013A&A...558A..17O, 2014MNRAS.443..947L}, and a ratio of $^{12}$CO/$^{13}$CO = 
25$\pm$2 has been determined by \citet{2015AJ....149...13M} from their high-resolution data. This 
enrichment in $^{13}$CO is an unmistakable sign for chemically processed matter around the central 
object \citep{2009A&A...494..253K} and hence excludes a pre-main sequence nature of MWC~137. An 
evolved, supergiant classification of MWC~137 was also postulated by \citet{1998MNRAS.298..185E} due 
to the high stellar luminosity, and is reinforced by \citet{2016A&A...585A..81M} based on the 
position of the star in optical and near-infrared color--magnitude and color--color diagrams. But 
despite the clear exclusion of a pre-main sequence nature, MWC~137 continues to appear in catalogs 
and studies about Herbig Be stars \citep[e.g.,][]{2017MNRAS.472..854A, 2018A&A...620A.128V, 
2019AJ....157..159A}, possibly owing to the star's peculiar characteristics.

A compact gaseous disk with a position angle (PA) of 120$\degr$ on the sky is inferred from H$\alpha$ 
spectro-polarimetric observations \citep{1999MNRAS.305..166O, 2002MNRAS.337..356V}. While the near- 
and mid-infrared colors point towards warm dust around MWC~137 
\citep{2016A&A...585A..81M, 2019Galax...7...83K}, there is a clear lack of cool dust in the close 
vicinity of the star \citep{2011ApJ...727...26S}. Moreover, the spectral energy distribution at 
millimeter and centimeter wavelengths is dominated by free-free emission \citep{2003ApJ...598L..39F}.

An optical high-velocity jet with several knots has been discovered as well 
\citep{2016A&A...585A..81M}, that seems to emanate from MWC~137. With a PA of 
18--20$\degr$, the jet is perpendicular to the postulated disk, but the origin of both jet and disk 
is still unclear. 

High-resolution K-band spectra of MWC~137 show molecular emission from CO with a kinematic broadening 
of 84$\pm$2\,km\,s$^{-1}$ projected to the line of sight \citep{2015AJ....149...13M}, and the 
emission has been interpreted as coming from a molecular gas ring, possibly revolving the star on a 
Keplerian orbit. Additional indication for rotating circumstellar gas is provided by the intense 
optical forbidden emission lines of [O\,{\sc i}] $\lambda\lambda$ 6300, 6364. 
\citet{2018MNRAS.480..320M} interpret the double-peaked profiles with broad wings of the [O\,{\sc i}] 
lines with a superposition of four distinct ring contributions having rotation velocities, projected 
to the line of sight, of 68.0, 46.8, 31.0, and 20.3\,km\,s$^{-1}$. Neither emission from [O\,{\sc 
i}]$\lambda$ 5577 nor from [Ca\,{\sc ii}] has been detected from MWC~137 that could help to further 
constrain the disk structure and its kinematics.

A highly debated issue is the distance to MWC~137. Former estimates spread from 1.3\,kpc 
\citep{1992ApJ...397..613H} up to $\sim 13$\,kpc \citep{1984ApJ...279..125F, 1988A&A...191..323W}.  
Based on the parallax measurement of $0.109624\pm 0.054560$\,mas provided by Gaia DR2, 
\citet{2020MNRAS.496.1051A} derived a distance of 13.6$^{+1.8}_{-1.3}$\,kpc whereas 
\citet{2018A&A...620A.128V} reported a value of 2.9074$^{+ 0.6048}_{-0.4011}$\,kpc using the same 
parallax. The Gaia Early Data Release 3 \citep{2016A&A...595A...1G, 2021A&A...649A...1G} provides a 
refined value of $0.19404\pm 0.02561$\,mas for the parallax, which is about 77\% higher than the 
previous one. Given the uncertainties in determining distances from measured parallaxes, we use the 
simple conversion with $d = 1/\mu$, where $\mu$ is the measured parallax, and obtain a value of 
5.15$^{+0.79}_{-0.60}$\,kpc. While we are aware that this conversion bears uncertainties 
\citep[see][]{2018A&A...616A...9L}, this value agrees reasonably well with the distance of 
$5.2\pm 1.4$\,kpc derived by \citet{2016A&A...585A..81M} from an analysis of the cluster stars within 
Sh~2-266. Therefore, we will adopt a distance of 5.15\,kpc to MWC~137.

In an attempt to determine the large-scale nebular kinematics \citet{2017AJ....154..186K} had obtained 
three long-slit spectroscopic observations covering H$\alpha$ as well as the nebular lines of 
[N\,{\sc ii}] and [S\,{\sc ii}]. Their analysis of the emission lines revealed a complex behavior of 
the radial velocities across the nebula. However the number of slit positions was insufficient to 
derive the global kinematics.

In this paper, we present new data sets that have been acquired to complement the previous 
observations with the aim to achieve a more complete prescription of the nebular kinematics. Our 
various data sets and their reduction procedures are introduced in Section~\ref{sect:obs}. In 
Section~\ref{sect:results} we present our analyses of the nebula expansion and kinematics, along with 
the computation of the electron densities across the nebula. In this section we also investigate the
photometric light curve of the central object. Our discussion on the nebula, stellar variability, and 
stellar parameters follow in Section~\ref{sect:discussion}, before we summarize our results in 
Section~\ref{sect:conclusions}.

\section{Observations and data reduction}\label{sect:obs}

\subsection{Optical long-slit spectroscopy}

Four long-slit optical spectra were obtained on 2018 November 8 with the Nordic Optical Telescope 
(NOT) using the Andalucia Faint Object Spectrograph and Camera (ALFOSC). ALFOSC's field of view (FOV) 
is $6\farcm 4\times6\farcm 4$ and the pixel scale is $0\farcs 21$ pix$^{-1}$. 

We utilized Grism~\#17 with a slit width of $0\farcs 5$, covering a spectral range of 6315-6760~\AA\ 
with a spectral reciprocal dispersion of 0.29~\AA\ pix$^{-1}$ ($R\sim 10\,000$). This choice is the 
same as for our previous observations \citep{2017AJ....154..186K}, covering H$\alpha$ and the nebular 
lines [N\,{\sc ii}] $\lambda\lambda$ 6548,6583 and [S\,{\sc ii}] $\lambda\lambda$ 6716,6731.

\begin{figure}
\begin{center}
\includegraphics[width=\hsize,angle=0]{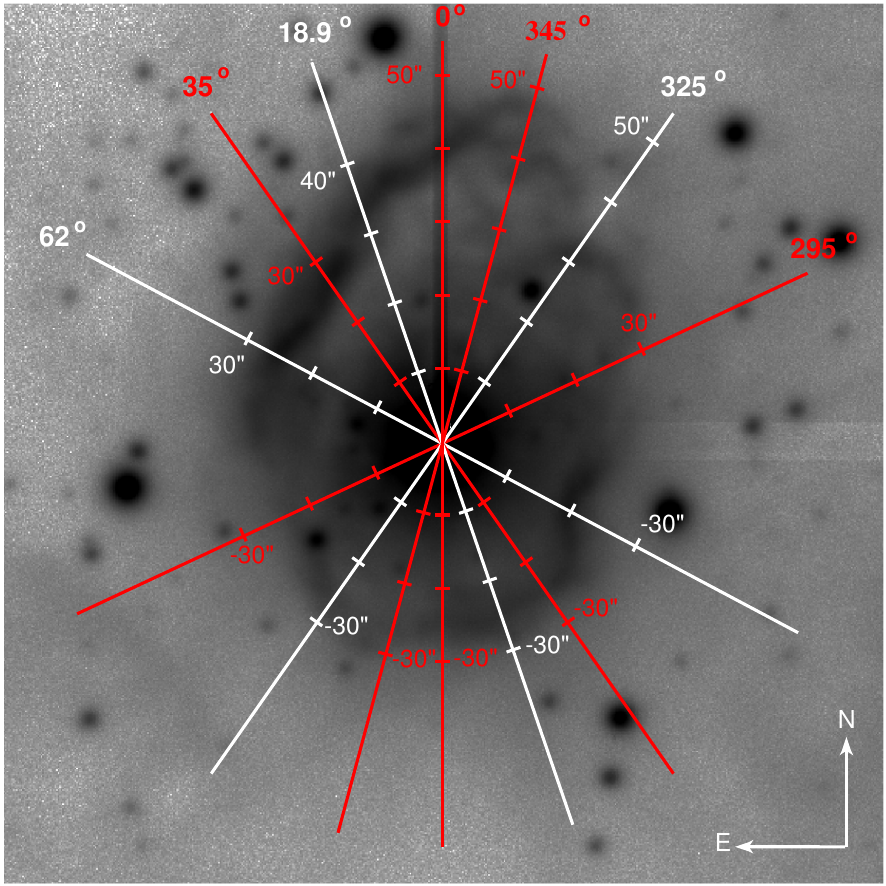}
\caption{ALFOSC H$\alpha$ image of the nebular structure around MWC\,137 from our previous observations 
in 2016 \citep{2017AJ....154..186K}. The FOV is $2\arcmin\times2\arcmin$. The new (red) slit 
positions have been included along with the old (white) ones. PAs are marked with boldface numbers. 
The distance scale in steps of $10\arcsec$ is marked with ticks along the slits in both directions 
from the central star.}
\label{fig:alfosc}
\end{center}
\end{figure}

The slits were centered on the star and had PAs of 0$\degr$, 35$\degr$, 295$\degr$, and 345$\degr$ 
(see Figure~\ref{fig:alfosc}) following the convention of measuring PAs from North to
East\footnote{Note that in \citet{2017AJ....154..186K} the PAs were accidentally measured from North 
to West.}. For each slit position the exposure time was 30 min, and the observations were carried 
out with a seeing of $\sim 1\arcsec$.

The spectra were reduced (bias, flat fielding, wavelength calibration) using standard 
IRAF\footnote{IRAF is distributed by the National Optical Astronomy Observatory, which is operated by 
the Association of Universities for Research in Astronomy (AURA) under cooperative agreement with the 
National Science Foundation.} routines, and were corrected for both the heliocentric and the systemic 
radial velocity (RV). For the latter we used the value of 42\,km\,s$^{-1}$ determined by 
\citet{2017AJ....154..186K}.

\subsection{Optical Fabry-Perot Interferometry}

Additional observations have been taken on 2019 November 19 and 21 with the Spectral Camera with 
Optical Reducer for Photometric and Interferometric Observations No 2 
\citep[SCORPIO-2,][]{2011BaltA..20..363A}, 
mounted at the Russian 6-m telescope (BTA). We used SCORPIO-2 in the Scanning Fabry-Perot mode in the 
order 751, operating with 40 spectral channels and providing a spectral resolution of 0.4\,\AA \ 
($R\sim 15\,400$). The total field of view of SCORPIO-2 is $6\farcm 1\times 6\farcm 1$. The CCD 
detector, E2V 42-90, has a size of 4600$\times$2048 pixels with a pixel size of 13.5$\times$13.5$\mu$m 
and a pixel scale of $0\farcs 18$ per pixel. During the observations with the FPI only the central part 
of the detector was used in $4\times4$ read-out binning mode providing the scale $0\farcs71$ per pixel.

The observations were centered on the [S\,{\sc ii}] $\lambda\lambda$ 6716,6731 lines.
We refrained from using the filter covering H$\alpha$ and the adjacent [N\,{\sc ii}] $\lambda$ 6583 
line because of the brightness of the central star in H$\alpha$ leading to strong saturation and 
blending effects when scanning the central regions. Instead, we gave preference to the two nebular 
lines [S\,{\sc ii}]. Their lower intensities allowed us to take longer exposures, hence deeper images, 
without the risk of saturation.

\begin{figure*}
\centering
\includegraphics[width=0.95\hsize]{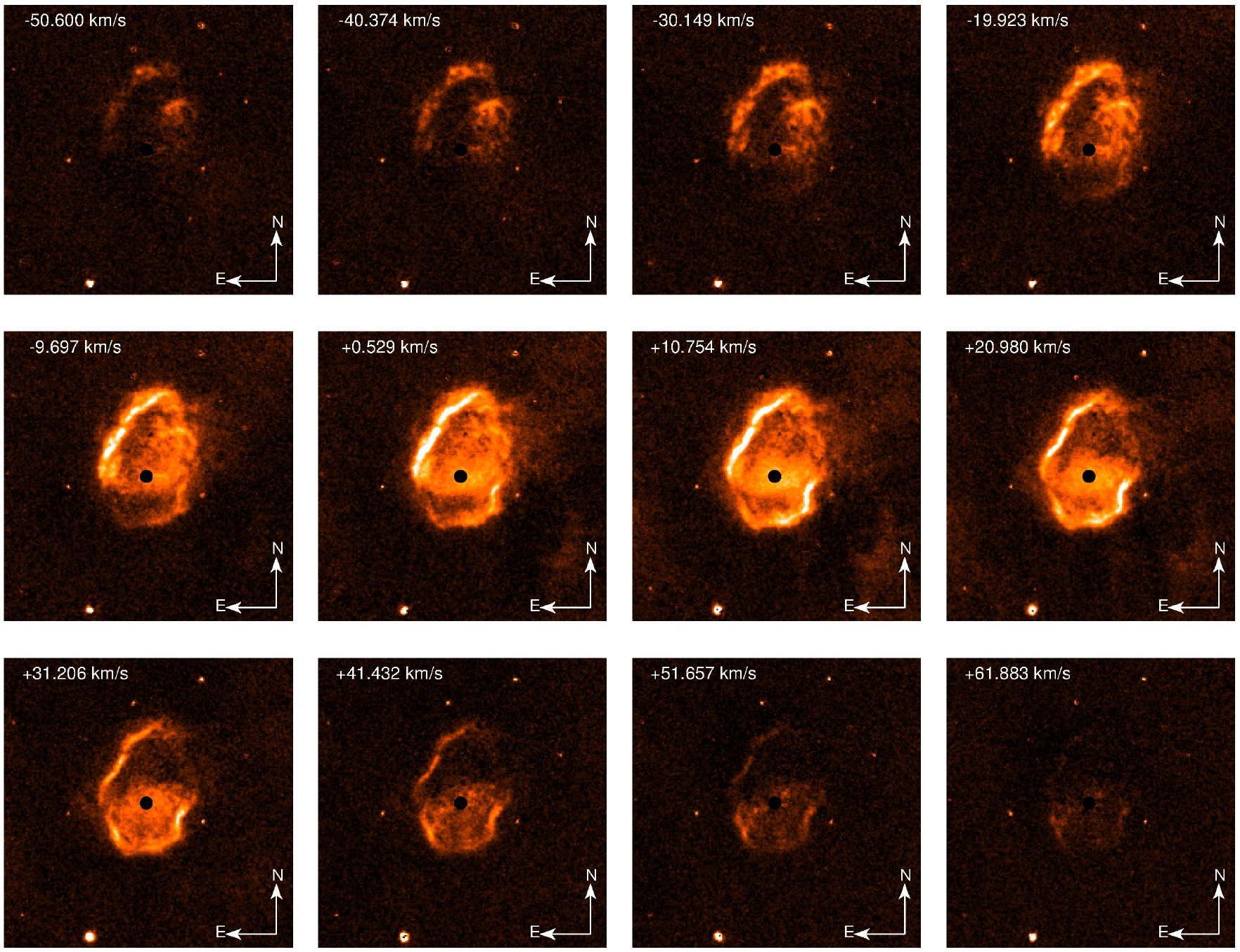}
\caption{FPI images of the co-added [S\,{\sc ii}]$\lambda\lambda$ 6716,6731 lines in various velocity 
bins. The image sizes are $3\arcmin \times 3\arcmin$ and have been centered on the stellar position. 
The central regions with a radius of 2.5$\arcsec$ have been masked. The intensity scale is linear and 
is the same in all images.} 
\label{fig:velocity_bins} 
\end{figure*}

During the first observing night, the observations were obtained in a single orientation, whereas two 
different orientations were used during the second night. The different orientations help to correct
for parasitic ghosts. For each set of observations, the exposure times were 30 sec per channel. Data 
reduction was performed following the recipes of \citet{2002BSAO...54...74M, 2015AstBu..70..494M} and 
\citet{2008AstBu..63..181M}. The reduced data cubes for the two orientations from the second night have 
been co-added. Unfortunately, the observing conditions were poor, with seeing changing from $2\farcs 
7-3\farcs 5$ to $3\farcs 1-3\farcs 8$ between the two observations. To combine the two data cubes, we 
had to degrade the one taken with the slightly better conditions, resulting in a final spatial 
resolution of just $3\farcs 8$. The data cube from the first night was taken under significantly 
better seeing conditions of $1\farcs 2-1\farcs 5$, resulting in a final spatial resolution of 
$1\farcs 5$.

The data have been corrected for heliocentric and systemic RV and continuum subtracted. 
Moreover, for each [S\,{\sc ii}] line the data have been converted into RV, and the 
emission in both lines has been co-added in velocity space for a better signal to noise ratio (SNR) of 
the nebular structure. The final SNR has values spreading from 5-10 in the faintest to 55-60 in the 
brightest nebular regions. 

\subsection{Optical imaging data}

A new H$\alpha$ image of the MWC 137 nebula was acquired with the 1.54-m Danish Telescope at La 
Silla on 2019 October 27. The Danish Faint Object Spectrograph and Camera (DFOSC) was exploited 
with a FOV of $13\farcm 7 \times 13\farcm7$ and a pixel scale of $0\farcs 39$\,pix$^{-1}$. The narrow 
band H$\alpha$ filter, ESO \#693, centered at 656.23\,nm and with a width of 6.21\,nm, was used. 
The total exposure time was 35 min, achieved with 7 dithered single exposures of 5 min each. 
The seeing of the combined frame was $1\farcs 5$.

In addition, we have at our disposal the raw images of MWC 137 presented by 
\citet{2008A&A...477..193M}. A single, 20 min long exposure was acquired on 2001 September 28 with 
the 60-inch telescope on Mt. Palomar with a seeing of $2\farcs 1$. The used FOV, pixel scale and 
narrow band H$\alpha$ filter are all comparable to our observations with the Danish Telescope. For 
more details on these observations see \cite{2008A&A...477..193M}. 

For both facilities, standard routines in IRAF were used for data reduction and for combining the
individual frames.

\subsection{Complementary data}

To date, MWC~137 (TIC 437994564) has been observed twice with high-cadence (120 s) by the Transiting 
Exoplanet Survey Satellite (TESS). The first observation period was between 2018 December 15 and 2019 
January 6 (BJD = 2458468.27--2458490.05) in Sector 6 with camera 1 and CCD 3. The second visit took 
place about two years later, between 2020 December 18 and 2021 January 13 (BJD = 
2459201.74--2459227.57) in Sector 33 with camera 1 and CCD 4.  We accessed the calibrated light 
curves\footnote{The data described here may be obtained from the MAST archive at 
{https://dx.doi.org/10.17909/t9-ncv5-bb52}.} through the Mikulski Archive for Space Telescopes (MAST). 
Calibrated means that the flux series has been corrected for common instrumental systematics, for the 
amount captured by the photometric aperture, and for crowding from known nearby stars. From these 
calibrated TESS light curves we used only the highest quality measurements for the analysis.

\section{Results}\label{sect:results}

\begin{figure*}
\begin{center}
\includegraphics[width=0.8\hsize]{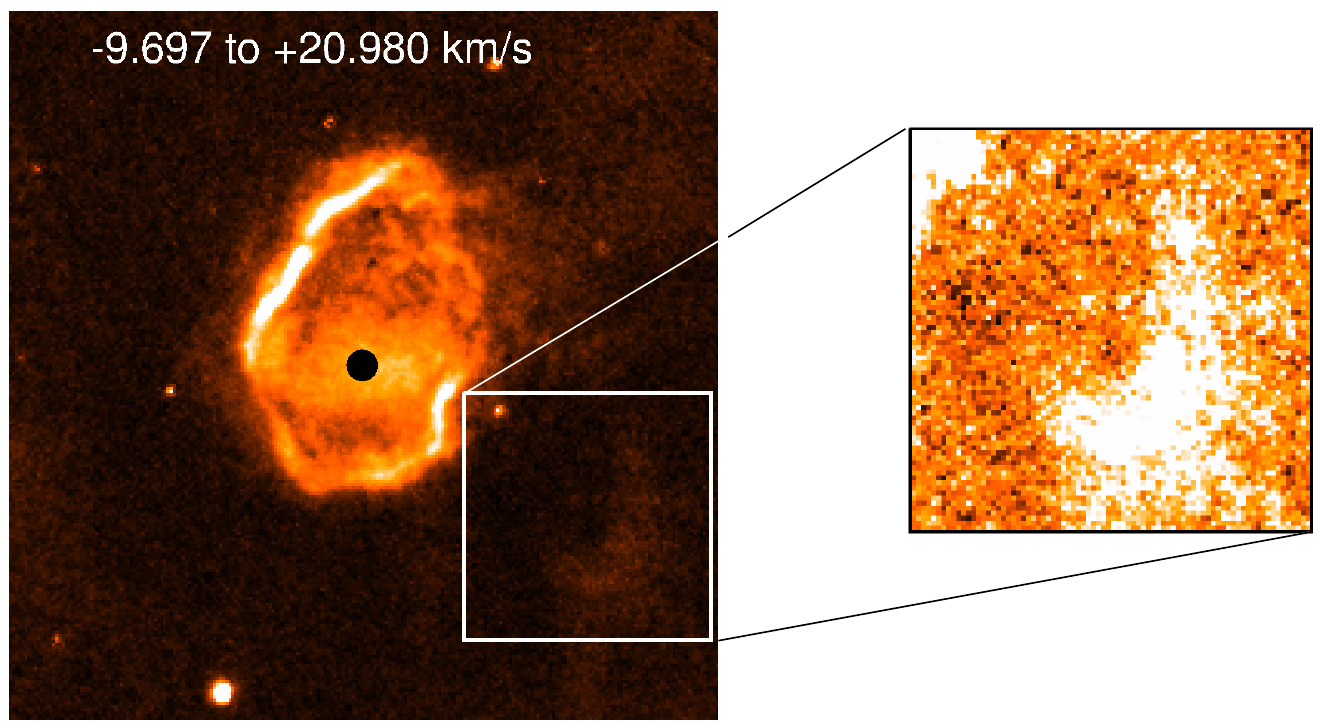}
\caption{Co-added [S\,{\sc ii}] emission in the velocity range $-$10\,km\,s$^{-1}$ to 
$+$21\,km\,s$^{-1}$. North is up and East is to the left. Ionized gas seems to emanate from the 
north-western nebular regime causing diffuse emission out to large distances, and a bow structure is 
detected in the south-western region. This bow is marked with the white box and magnified in the right 
panel using a higher intensity contrast.}
\label{fig:bow} 
\end{center}
\end{figure*}

\subsection{Possible expansion in the plane of the sky}

Our new image of MWC~137 from 2019 was used in comparison to the image from 2001 to explore any 
morphological changes and/or expansion of the large scale nebula in the plane of the sky over the 
total time span of 18.1 years. We utilized more than 40 stars in the FOV to first match both images, 
pixel by pixel, using the IRAF tasks {\it geomap} and {\it geotran}. The RMS of the matching was 0.1 
pixels. From a first glance, no obvious morphological evolution or expansion of the nebula was seen 
from a simple by-eye inspection (blinking) of the images. 

Next, we used the so-called \textit{magnification method} \citep[see e.g.][]{1999AJ....118.2430R, 
2018A&A...612A.118L} which uses the residual images to detect displacements in the nebula down to 0.1 
pixels. For this, further image processing was required, such as matching the seeing of the 2019 
image to the slightly worse one of the 2001 image, and matching the flux. This matching was performed 
as described in Section 3 and 4.1 in \citet{2018A&A...612A.118L}. 

After that, the 2001 frame was magnified, using the IRAF task {\it geotran} with a steadily 
increasing magnification factor, and then subtracted from the 2019 image. If the nebula would have 
expanded in the plane of the sky, the resulting residual images should reveal a distinct pattern 
based on which the magnification could be determined \citep[see][]{1999AJ....118.2430R}. However, our 
analysis revealed the we detected neither any morphological changes nor an expansion in the plane of 
the sky. 

This means that, considering the time span of 18.1 years, a possible expansion in the plane of the 
sky is proceeding with a tangential velocity of less than 52.7\,km\,s$^{-1}$ for the pixel scale of 
$0\farcs 39$ and the distance of 5.15\,kpc towards MWC~137. In fact, the maximum RV 
component measured in the nebula is only about 40\,km\,s$^{-1}$ \citep[see][and Section 
\ref{Sect:longslit} below]{2017AJ....154..186K} and hence too small to create detectable changes
in the nebular morphology to date. For the idealized case of a ballistically expanding sphere with a 
tangential velocity of 40\,km\,s$^{-1}$, we can hence expect to detect an indication of expansion
earliest after a time span of about 24 years, that is in the year 2025. However, the lack of 
significant RV of the nebula around the position of the star questions a ballistic 
expansion so that the real expansion of the nebula in the plane of the sky might proceed (much) 
slower.

\subsection{The FPI images}\label{kinematics}

Although the data from the second night had longer exposure times, we present the data from the first
night because of their significantly higher spatial resolution. The lack of a second orientation 
observation resulted in several artefacts that appear on the final cube, such as false emission at both 
the stellar position and the position of several stars in the vicinity. These could not be properly 
corrected but were identified based on the deeper data cube. Within the nebula structure, only at the 
central star's position a ghost artifact was identified and masked and is excluded from our discussion. 
This masked area around MWC~137 has a radius of 2.5$\arcsec$, corresponding to a sphere with a radius 
of 0.062\,pc.

Slicing the FPI data cube with respect to RV, we find that the nebula is detected from about 
$-$51\,km\,s$^{-1}$ to about $+$62\,km\,s$^{-1}$ with respect to the nebular systemic velocity, as 
shown in Figure~\ref{fig:velocity_bins}. For a better comparison, the intensity in all velocity bins 
is displayed on a linear scale, and in all panels the intensity range is the same. 

Starting from the very blue, the first features that become visible are two structures in the 
northern nebular regions. The north-eastern ring or arm-like structure brightens until it reaches 
maximum intensity at $+$0.5\,km\,s$^{-1}$ and then starts to fade and disappear for velocities redder 
than $+$52\,km\,s$^{-1}$. The second, most blue-shifted emission is related to a smaller ring or 
arm-like structure north-west of MWC~137. The northern domain of this feature reaches maximum intensity 
at $-$20\,km\,s$^{-1}$ and disappears for velocities redder than $+$11\,km\,s$^{-1}$. Starting from 
about $-$30\,km\,s$^{-1}$, a south-western arm starts to appear. Its maximum intensity is reached 
around $+$11\,km\,s$^{-1}$, and it remains visible beyond red-shifted velocities of 
$+$52\,km\,s$^{-1}$. Around $+$11\,km\,s$^{-1}$, this feature seems to connect to the large 
north-eastern arm structure. A further, high intensity region appears west of the central object 
in the velocity interval $-$10\,km\,s$^{-1}$ to $+$20\,km\,s$^{-1}$. In the velocity bins at
$+0.5$\,km\,s$^{-1}$ and $+$11\,km\,s$^{-1}$, this emission seems to spread around the central object 
in east-west direction. Finally, the south-western arm-like structure embraces some diffuse emission 
in the southern region with velocities from $+$31\,km\,s$^{-1}$ to approximately 
$+$52\,km\,s$^{-1}$.

Furthermore, the large FOV of the FPI reveals emission features at greater distances from the star, in 
regions which have not been covered by the MUSE mosaic of \citet{2016A&A...585A..81M}. These are (i) 
diffuse emission apparently emanating from the north-western nebular regime, and (ii) a bow structure 
towards the south-west at a PA of 225--245$\degr$. Both emission structures are visible on our 
H$\alpha$ images taken with ALFOSC and with the Danish telescope, and are also present on the image of 
\citet{2008A&A...477..193M}. But \citet{2008A&A...477..193M} mentioned only the diffuse emission. 
The presence of the bow structure on all images taken between 2001 and 2019 and with different 
instruments implies that it is real and not an artifact.

\begin{figure*}
\centering
\includegraphics[width=0.49\hsize]{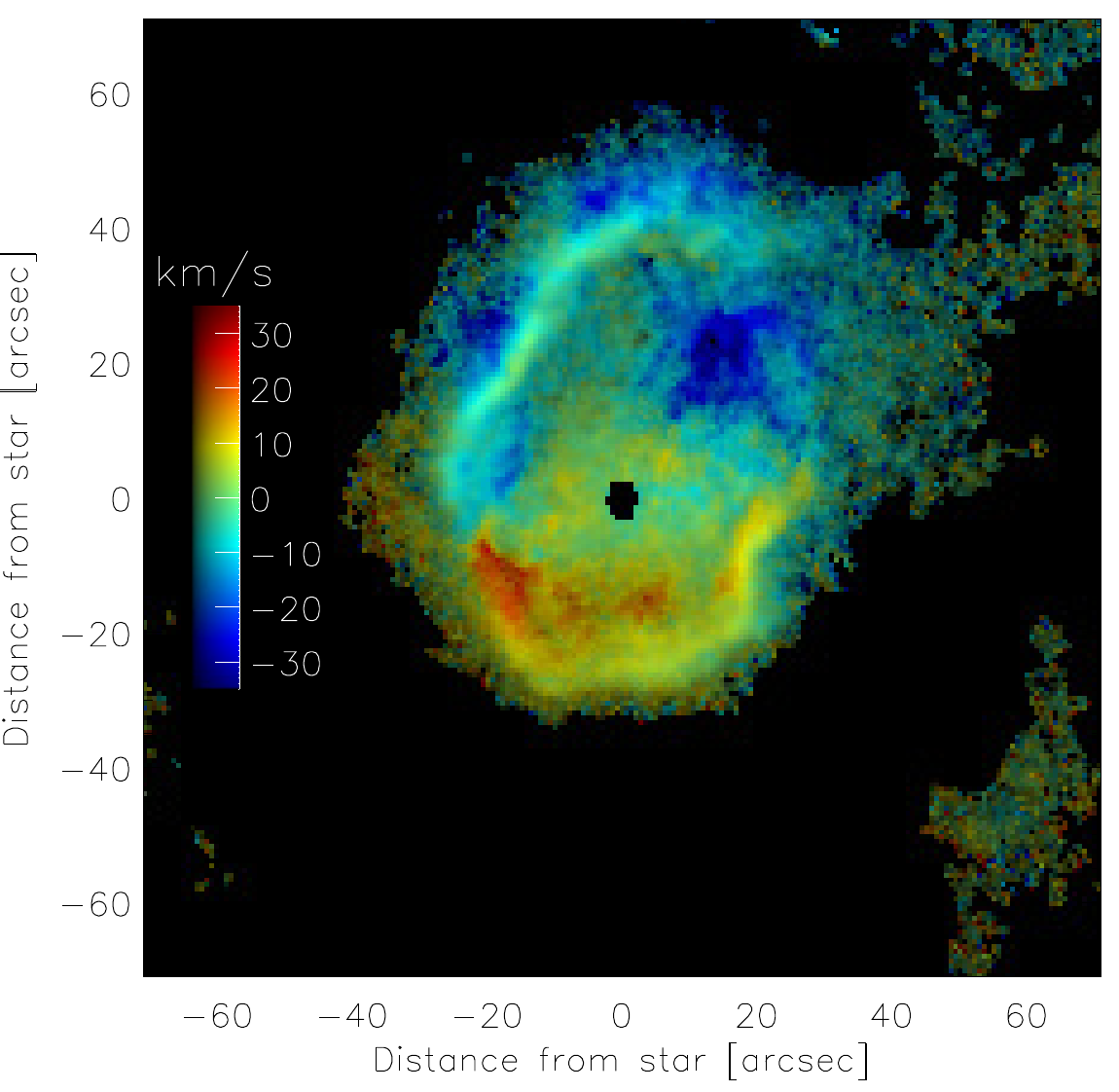}
\includegraphics[width=0.49\hsize]{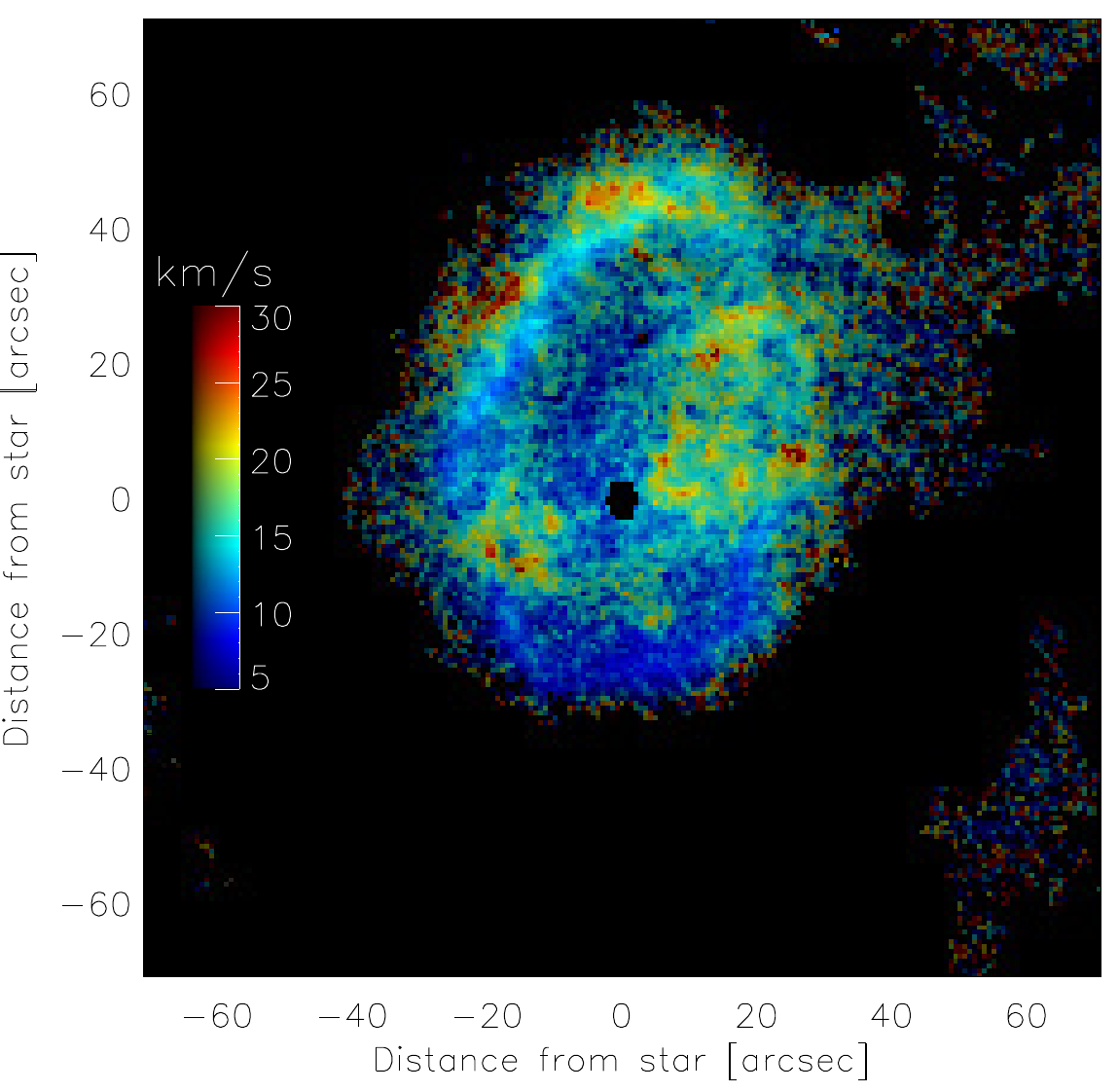}
\caption{FPI images of the [S\,{\sc ii}] emission. North is up and East is to the left.
In color are shown the radial velocity distribution (left panel) and the velocity dispersion (right 
panel) corrected for instrumental broadening. The pixel brightness is scaled according to the 
[S\,{\sc ii}] line intensity.}
\label{fig:velocity_dispersion} 
\end{figure*}

These two emission features appear in the velocity interval from $-$10\,km\,s$^{-1}$ to 
$+$21\,km\,s$^{-1}$. For better visibility of these features, we co-added the images in this 
velocity range and show the combined image in Figure~\ref{fig:bow}. The vertex of this bow or arch 
has a distance of about 80$\arcsec$ from the central star and an extension of about 36$\arcsec$ from 
north-west to south-east. These measurements translate into physical scales, projected to the plane of 
the sky, of 2\,pc and 0.9\,pc. But whether its origin is linked to MWC~137 is difficult to tell. At 
least, this arch or bow seems not to be related with the jet discovered by \citet{2016A&A...585A..81M}, 
which emerges at a PA of 18--20$\degr$. 

Finally, we use the FPI data cube to measure the radial velocities and velocity dispersions of the 
combined [S\,{\sc ii}] lines over the full nebula. Visual inspection of the profiles reveals that the 
lines are single-peaked everywhere in the nebula. The emission lines are fitted with a Voigt profile, 
and the velocity dispersion has been corrected for instrumental broadening as described in 
\citet{2008AstBu..63..181M}. The results are shown in Figure~\ref{fig:velocity_dispersion}. The general 
trend is that the intense nebula arm structures have lower red- or blue-shifted velocity compared to 
the gas inside and outside (in projection onto the plane of the sky) of these structures. In addition, 
we observe that the regions with highest radial velocities have the tendency to also display high 
velocity dispersion, but with no strict coincidence. 

The choice of the filter setting for these observations was optimized such that both [S\,{\sc ii}] 
lines were simultaneously observed. However, as we can see in Figure~\ref{fig:velocity_dispersion}, the 
emission can become rather broad in some nebular regions, so that not always the full profile of either 
the blue or the red [S\,{\sc ii}] line was covered by the narrow filter, hampering a more detailed 
analysis of the physical conditions across the entire nebula.

\subsection{Radial velocities along the long-slit spectra}\label{Sect:longslit}

The long-slit spectra taken with ALFOSC cover, besides H$\alpha$, the forbidden emission lines 
[N\,{\sc ii}] $\lambda\lambda$ 6548,6583 and [S\,{\sc ii}] $\lambda\lambda$ 6716,6731. The 
slightly better observing conditions under which these spectra were taken, compared to the 
FPI images and, in particular, the longer spectral coverage allow for more precise measurements of the 
kinematics, as well as a proper determination of the physical parameters of the nebula. 

\begin{figure*}
\begin{center}
\includegraphics[width=\hsize,angle=0]{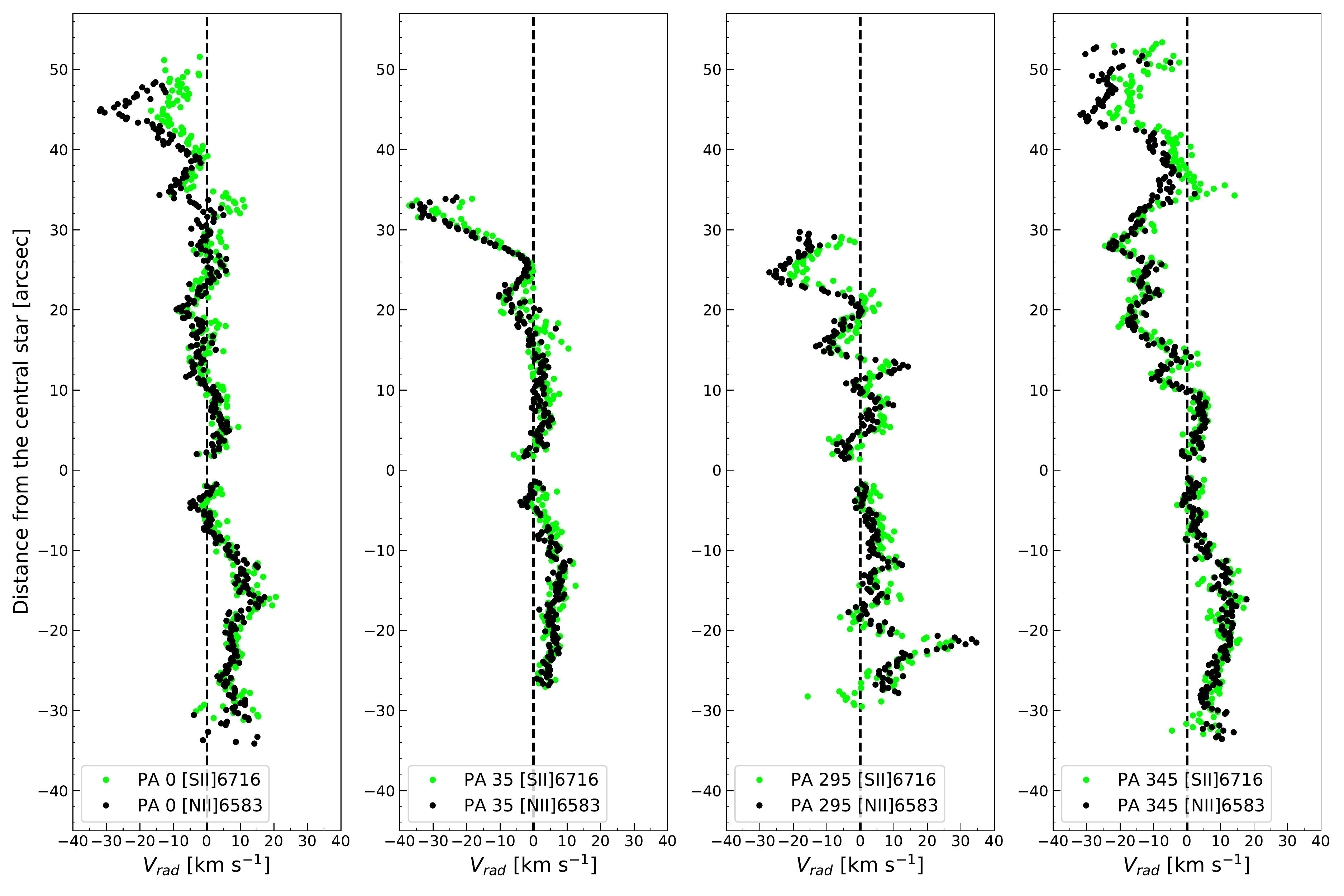}
\caption{Radial velocities of the [N\,{\sc ii}] 
6583~\AA\ (black) and [S\,{\sc ii}] 6716~\AA\ (green) emission lines. Positive distance from the 
central star on the image is towards the North, negative towards the South.}
\label{fig:rv-new} 
\end{center}
\end{figure*}

For both doublets we focused on their more intense line to perform the RV measurements. These are 
[N\,{\sc ii}] $\lambda$ 6583 and, over large regions of the nebula, [S\,{\sc ii}] $\lambda$ 6716. For 
each line, we fit the emission profiles in the 2D spectra, line by line, to determine RV as a 
function of distance from the star. The lines appear single-peaked and are not blended so that we fit 
them with a single Gaussian profile. These measurements are presented for all four 
PAs in Figure~\ref{fig:rv-new} and are provided in an on-line table following the layout of 
Table~\ref{tab:rv}. Profiles were excluded from the measurements if they were (i) too weak, (ii) 
contaminated by a cosmic ray that could not be removed, or (iii) contaminated 
with a neighboring line. The latter happens especially for the [N\,{\sc ii}] $\lambda$ 6583 line in 
the vicinity of the central star where H$\alpha$ is saturated, preventing any precise measurements of 
that line. The uncertainties of the velocity measurements are rather small, with an average value of 
0.95\,km\,s$^{-1}$, and are included in Table~\ref{tab:rv}. They have been determined from the RMS of 
the wavelength calibration. Contributions from the fitting to the total error are negligible. 

From Figure~\ref{fig:rv-new} we can see that in general both nebular lines display the same trend in
the gas dynamics. However, some deviations are noticeable. 
First, in the close vicinity of the hot central star the [S\,{\sc ii}] lines disappear and the 
[N\,{\sc ii}] lines become very weak on all our spectra. This points towards a higher electron density 
favoring collisional depopulation of the levels, along with a hotter gas and, therefore, preferentially 
higher ionized material. In fact, the higher density is obvious from the much more intense H$\alpha$ 
emission towards the center, whereas the slightly higher ionization state follows from high-resolution 
spectra of the innermost 2$\arcsec$ that display considerable emission of [S\,{\sc iii}] $\lambda$ 6312 
and only very weak emission in [S\,{\sc ii}] $\lambda$ 6716 
\citep[][their Figure~3]{2017AJ....154..186K}. These hindrances in measurements result in a gap in the 
RV values in all PAs within a radius of about 1$\farcs$5--2$\farcs$0 from the central star.

\begin{deluxetable}{llrcr}
\tabletypesize{\scriptsize}
\tablecaption{Radial Velocity Measurements \label{tab:rv}}
\tablewidth{0pt}
\tablehead{
\colhead{PA}   & \colhead{$\lambda_{\rm lab}$} & \colhead{RV} & \colhead{RV uncertainty} &\colhead{Dist}         \\
\colhead{($\degr$)}  &  \colhead{(\AA)}   &  \colhead{(km~s$^{-1}$)}   & \colhead{(km~s$^{-1}$)} 
&\colhead{($\arcsec$)}
}
\startdata
0 & 6716.440  & 10.90  & 0.94  & $-$31.15 \\
0 & 6716.440  &  2.02  & 0.94  & $-$30.94 \\
0 & 6716.440  & 15.32  & 0.94  & $-$30.73 \\
0 & 6716.440  & 15.05  & 0.94  & $-$30.52 \\
0 & 6716.440  &  7.91  & 0.94  & $-$30.31 \\
\enddata
\tablecomments{The entire table is published only in the electronic
  edition of the article.  The first 5 lines are shown here for
  guidance regarding its form and content.}
\end{deluxetable}

\begin{figure*}
\begin{center}
\includegraphics[width=0.7\hsize]{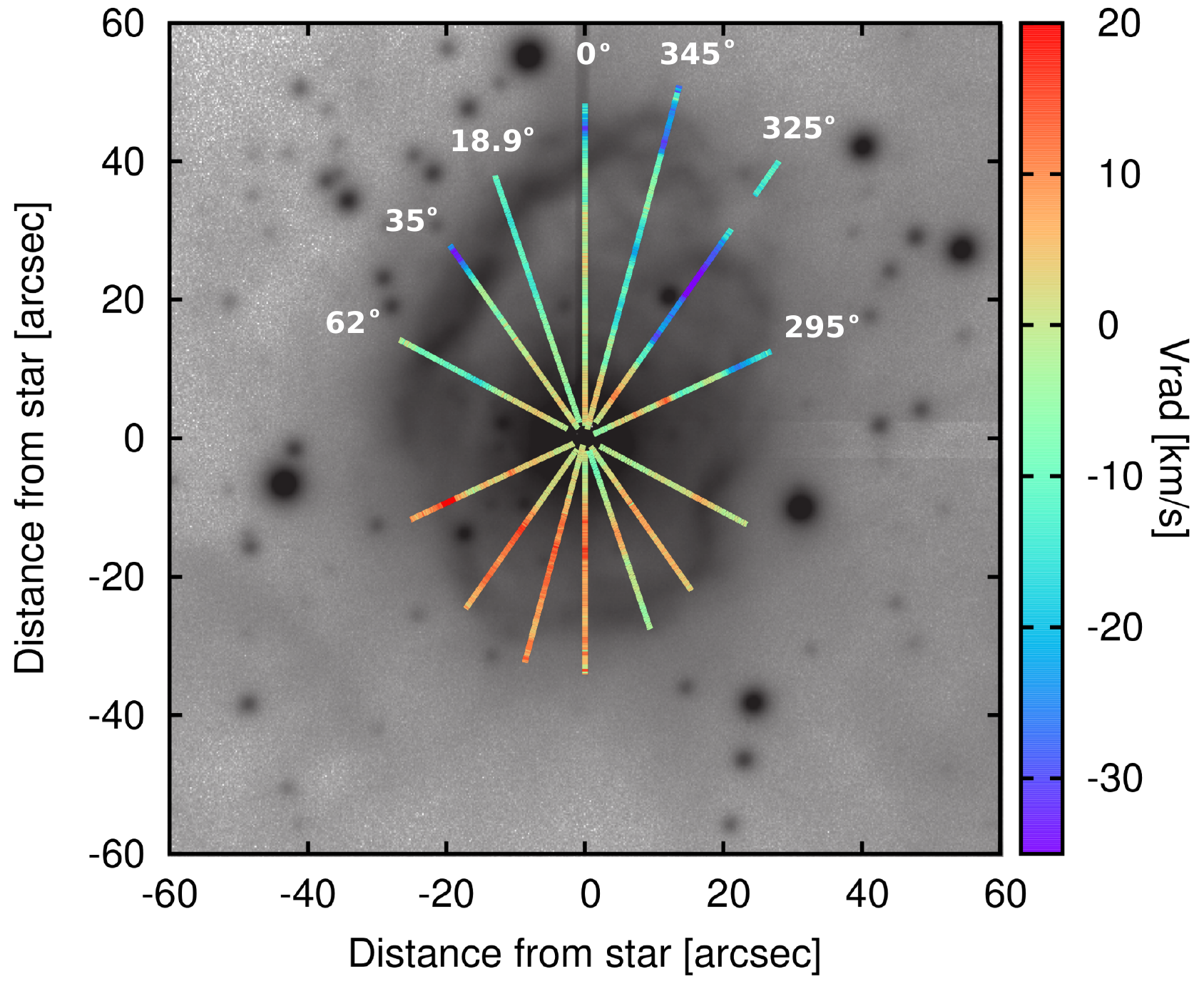}
\caption{Color presentation of the radial velocities along the different slits measured from the
[N\,{\sc ii}] 6583~\AA\ line.}
\label{fig:rv-color} 
\end{center}
\end{figure*}

Second, we notice considerable differences in the RVs of the two lines in the regions with highest 
blue-shifted RV values towards to north and north-west for PAs at $0\degr$, $325\degr$, and 
$295\degr$ \citep[the same is seen for PA $345\degr$,][]{2017AJ....154..186K} where the [N\,{\sc ii}] 
displays significantly higher velocities than the [S\,{\sc ii}]. We see the same trend in the 
red-shifted RV maximum for PA = $295\degr$ in the south-east. The deviations can reach values of 
$\Delta v_{\rm rad} \approx 10-15$\,km\,s$^{-1}$. Because these differences are significantly larger 
than the measurement errors, they are considered as real. Taking into account that the ionization 
potential of N\,{\sc i} (I.P.  = 14.5\,eV) is higher than that of S\,{\sc i} (I.P. = 10.4\,eV), and 
that the ionization of the elements is governed by the radiation flux from the hot central star, the 
emission in  [N\,{\sc ii}] should be formed closer to the star than the emission in [S\,{\sc ii}]. A 
relation between RV and the ionization potentials of the elements has been found in 
H\,{\sc ii} regions \citep[see, e.g.,][]{1999A&A...349..276E, 2019Ap.....62...57C} where it has been 
used to trace the velocity stratification along the line of sight. Application to the nebula of 
MWC~137 suggests that within these specific regions we see deceleration of the material from inside 
out.

\begin{figure*}
\begin{center}
\includegraphics[width=\hsize,angle=0]{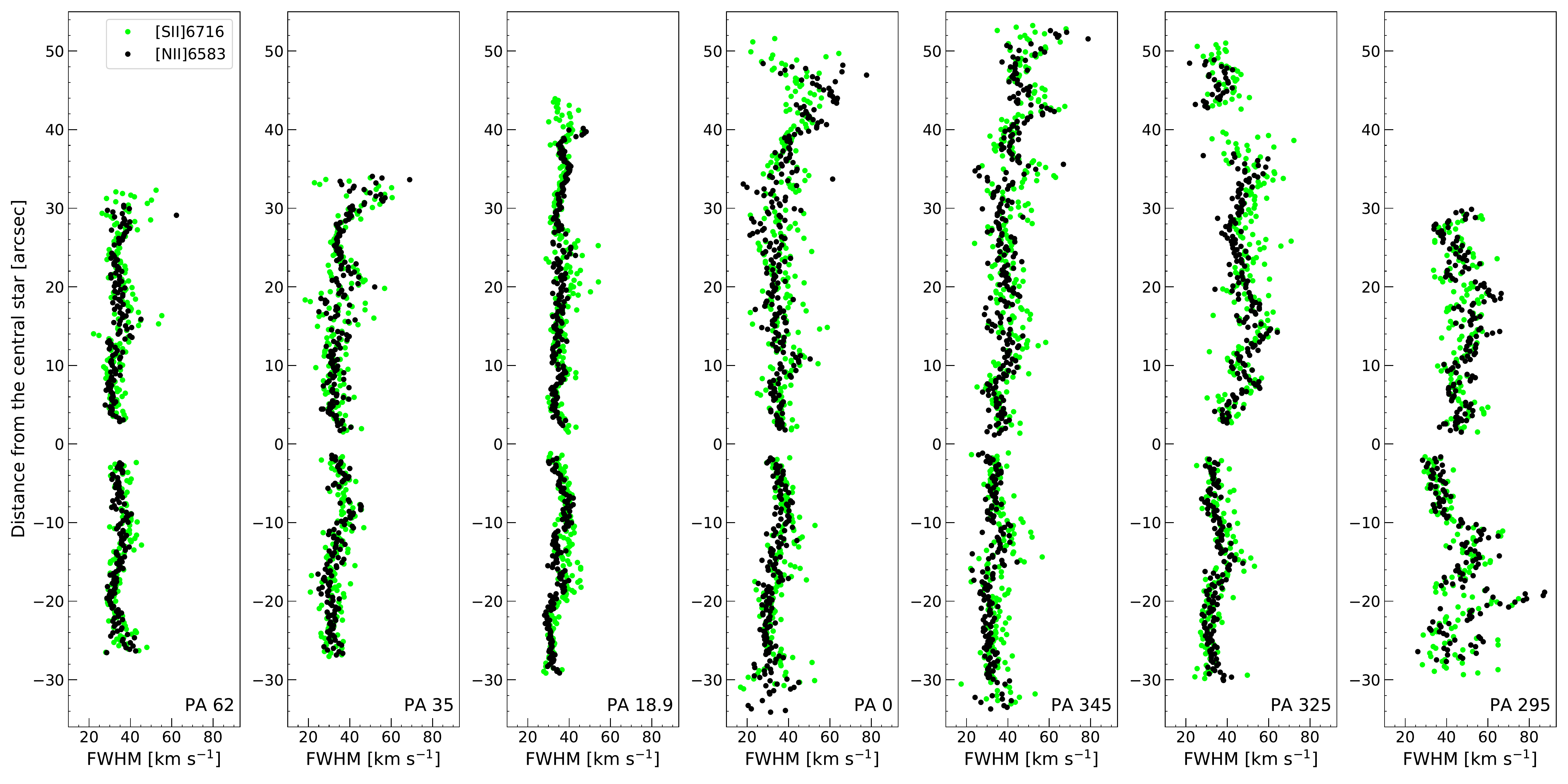}
\caption{Measured FWHM values of the [N\,{\sc ii}] 6583~\AA\ (black) and [S\,{\sc ii}] 6716~\AA\ 
(green) emission lines along all seven PAs. Positive distance from the central star on the image is 
towards the North, negative towards the South.}
\label{fig:fwhm} 
\end{center}
\end{figure*}

In agreement with the FPI data, the long-slit spectra reveal that the northern nebular region is 
predominantly blue-shifted with respect to the nebular systemic velocity, reaching values up to 
$-$36.2\,km\,s$^{-1}$, whereas the southern part is mostly red-shifted with velocities up to 
$+$34.6\,km\,s$^{-1}$. But despite this general trend, we also note strong variability of the RV values 
along most PAs, with multiple maxima and minima, in agreement with the inhomogeneous velocity pattern 
across the nebula seen in Figure~\ref{fig:velocity_dispersion} and with the wiggly structures 
detected along the PAs presented by \citet{2017AJ....154..186K}.

To visualize the RVs over the measured nebular regions, we utilize the RV values of 
the [N\,{\sc ii}] $\lambda$ 6583 line and combine them with those of our previous long-slit spectra 
for the three PAs at 18.9$\degr$, 62$\degr$, and 325$\degr$ \citep[respectively at 341.1$\degr$, 
298$\degr$, and 35$\degr$ according to the notation in Figure~4 of][]{2017AJ....154..186K}. In 
Figure~\ref{fig:rv-color} we present the color-coded RVs along all seven slits. For a comparison 
with the nebular structures, these velocities are overlaid on the H$\alpha$ emission nebula.  

From this image we can deduce that the largest region displaying highly blue-shifted RVs with values 
exceeding $-$40\,km\,s$^{-1}$ resides in the north-western area of the nebula, in particular along PA 
$=$ 325$\degr$. This area corresponds to the region in which neither cold molecular gas nor 
significant emission from warm dust is seen \citep{2017AJ....154..186K}. A further, though smaller 
blue-shifted domain occurs towards the north-east, beyond the outer rim (maximum intensity in the 
H$\alpha$ image) of the nebula at PA $=$ 35$\degr$. Here, the velocities reach values around 
$-$36\,km\,s$^{-1}$ whereas they tend towards zero within the region of maximum intensity. A similar 
behavior is found for the most northern region along PA $=$ 0$\degr$. Also here the RV is around zero 
on top of the maximum intensity and reaches highest blue-shifted values beyond that region. For the 
other PAs with blue-shifted emission (295$\degr$, 325$\degr$, and 345$\degr$) the situation is 
different. Here, the emission with maximum blue-shift either coincides with the intensity peaks or 
occurs at distances closer to the central star.

Turning to the southern nebular region we find the highest red-shifted velocities in the south-eastern
portion, with a narrow peak of about $+$35\,km\,s$^{-1}$ reached within an apparent gap in the 
H$\alpha$ double-ring structure along PA 295$\degr$. For the PAs at 325$\degr$, 345$\degr$, 
and 0$\degr$, the highest red-shifts occur (much) closer to the central star and are not correlated 
with the regions of maximum emission intensity. The only exception is PA $=$ 62$\degr$. Here the 
red-shifted emission coincides with the maximum intensity. But the red-shift in that region is with 
values smaller than $+$10\,km\,s$^{-1}$ much less prominent than along other PAs.

Our measurements also provide the FWHM values of the emission lines. These are shown in 
Figure~\ref{fig:fwhm} for [N\,{\sc ii}] 6583~\AA \ and [S\,{\sc ii}] 6716~\AA \ along all seven PAs. 
As for the RV values, we find that both forbidden lines display the same behavior of their FWHM 
values, with noticeable differences only in regions in which the lines are very faint and thus have a 
higher measurement uncertainty. For most PAs, a significant portion of the nebula displays some 
average FWHM value of 30--40\,km\,s$^{-1}$ (whereby 30\,km\,s$^{-1}$ correspond to the velocity 
resolution of the long-slit spectra), but in certain regions it can reach values of 
70--80\,km\,s$^{-1}$. Exceptions are the PAs at 325$\degr$ and 295$\degr$, which exhibit considerably 
higher FWHM values over large nebular portions. We also find that the nebular areas with high FWHM 
values correlate more or less with the regions of high RV, in agreement with what is seen in 
Figure~\ref{fig:velocity_dispersion}.

\subsection{Electron densities}

From their low-resolution MUSE data \citet{2016A&A...585A..81M} estimated average values of 
$T_{e}\sim 10\,000$\,K and $n_{e}\sim 300$\,cm$^{-3}$ for the electron temperature and density across 
the nebula. Our spectra have three times higher resolution, but are short in wavelength coverage and 
thus contain just two strategic forbidden lines, [S\,{\sc ii}] $\lambda\lambda$ 6716,6731, which allow 
us to investigate the electron density distribution along the PAs, but we cannot estimate the electron 
temperature. 

\begin{figure}
\centering
\includegraphics[width=\hsize,angle=0]{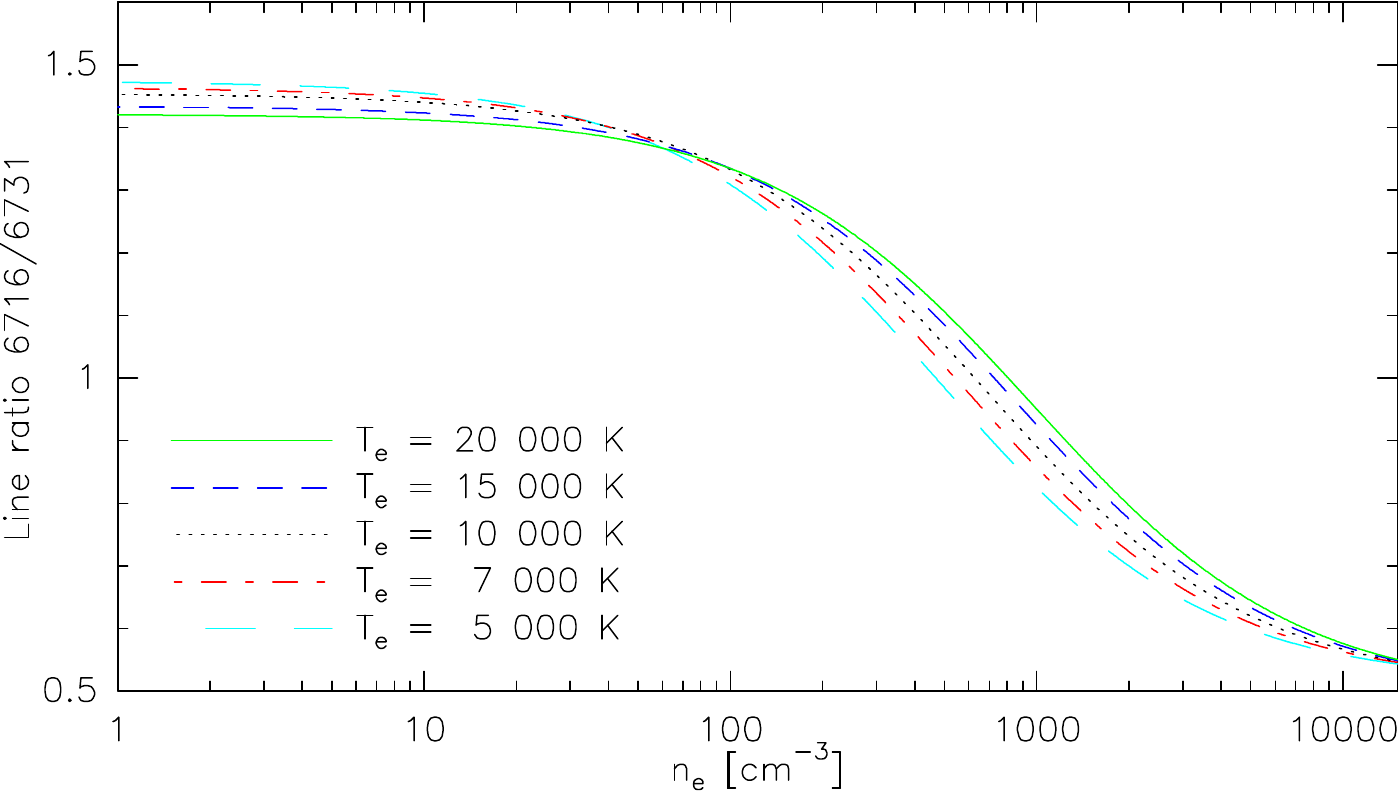}

\smallskip
\includegraphics[width=\hsize,angle=0]{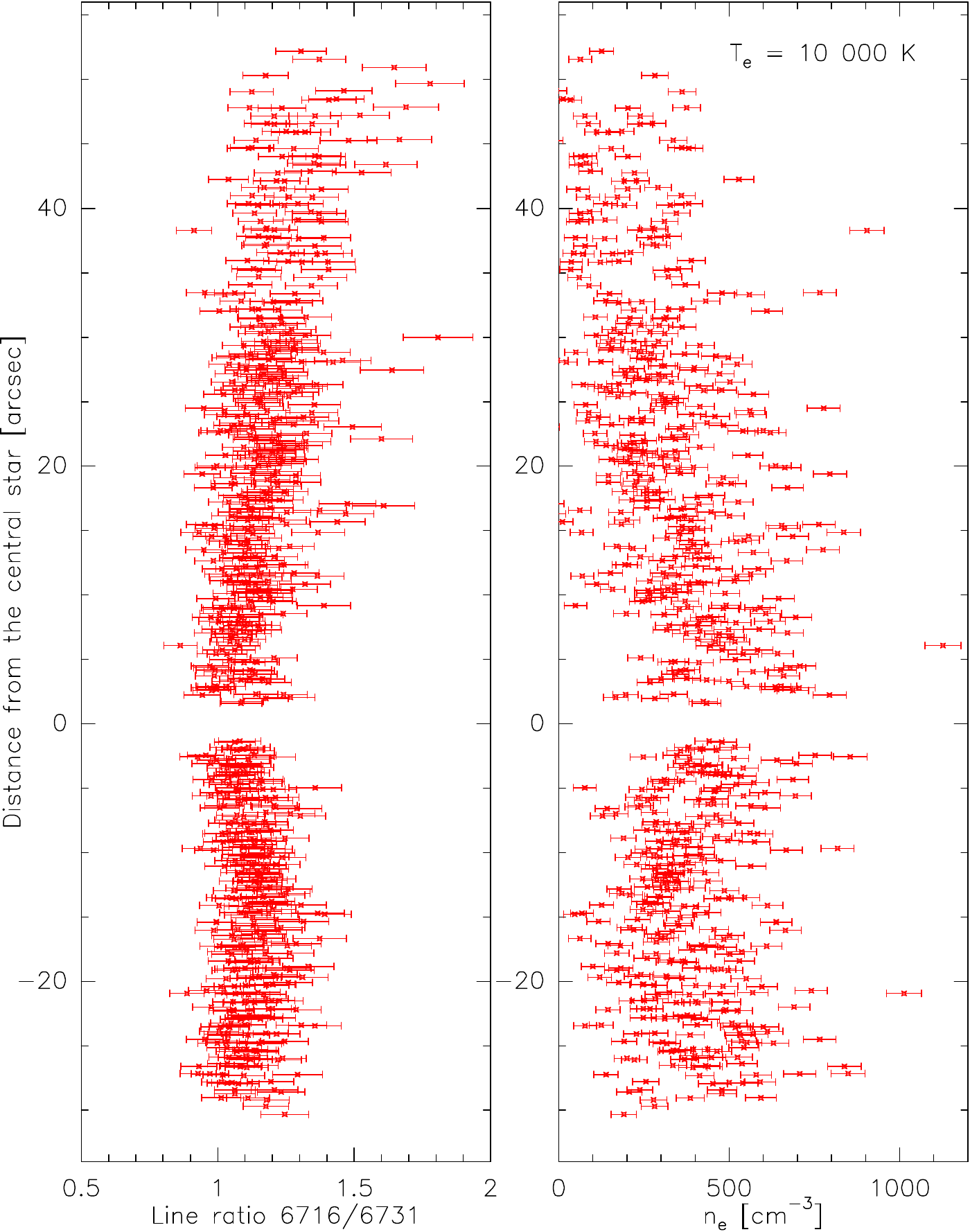}
\caption{Top: theoretical line ratios as function of electron density and for different electron 
temperatures. Bottom: Observed line ratios (left) for the measurements along all PAs, and derived 
electron densities (right) for a nebular temperature of 10\,000\,K.}
\label{fig:ratio_dens} 
\end{figure}

From our spectra, we measure, line by line, the line fluxes of both [S\,{\sc ii}] lines and then 
average the values every three lines before computing their ratio. This binning reduces the 
measurement errors, particularly in the very faint nebular regions, and also represents better the real 
values considering a seeing of $\sim 1\arcsec$ during the observations compared to the pixel size of 
$0\farcs21$. A conservative uncertainty estimate of the flux measurements is 5$\%$ resulting in an 
uncertainty on the observed line ratios of about 7$\%$. 

From theory, the intensity ratio of these two [S\,{\sc ii}] lines is given by
\begin{equation}
\frac{I(6716)}{I(6731)} = \frac{N(6716) A(6716) \lambda(6731)}{N(6731) A(6731)\lambda(6716)}\, ,
\end{equation}
where the parameters $N$, $A$, and $\lambda$ denote the level population, Einstein transition 
probability, and laboratory wavelength of the corresponding transition. This ratio is an implicit 
function of the electron density for a given temperature and vice versa. 

To compute the level population we consider a five-level atom in which the levels, from which the 
forbidden lines arise, are excited by collisions with free electrons. Collision parameters and 
Einstein transition probabilities are taken from \citet{2010ApJS..188...32T}. The level population 
is calculated for electron temperatures between 5\,000\,K and 20\,000\,K and for electron 
densities ranging from 1 to 15\,000\,cm$^{-3}$. The line ratios as a function of electron 
density are shown in the top panel of Figure~\ref{fig:ratio_dens} for five different electron 
temperature values, whereas the observed line ratios from North to South across the nebula are
plotted in the bottom left panel, and the resulting electron densities, considering a constant 
electron temperature of the nebula of 10\,000\,K, in the bottom right panel. 

\begin{figure*}
\begin{center}
\includegraphics[width=0.7\hsize]{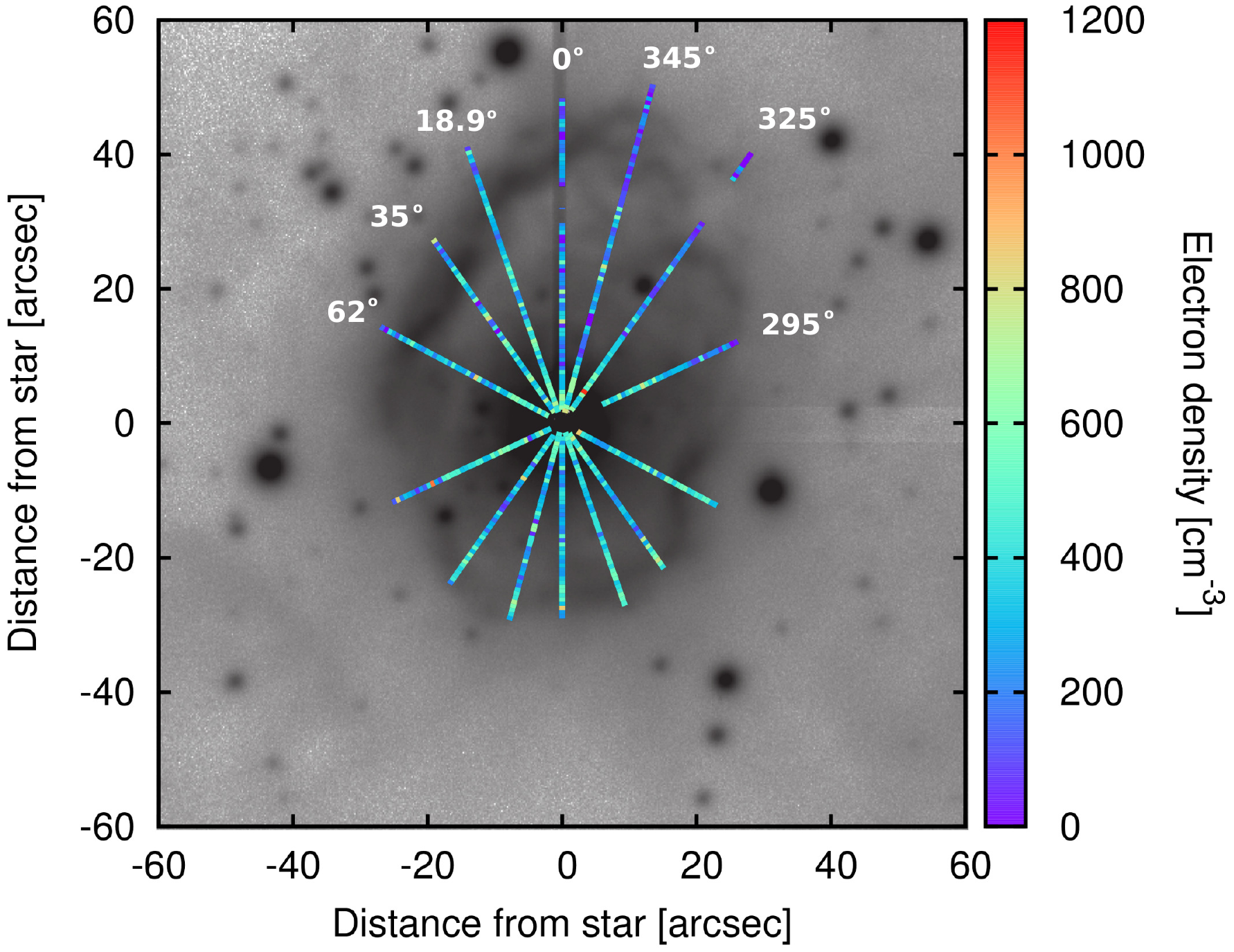}
\caption{Color presentation of the electron density along the different slits for a constant electron temperature of 10\,000\,K across the nebula.}
\label{fig:eldens-color}
\end{center}
\end{figure*}

The densities display high local fluctuations with the majority of measurements spreading from 200 to 
500\,cm$^{-3}$. But we also find a clear increase in electron density towards the position of the 
central star. This can be seen in the color-coded image shown in Figure~\ref{fig:eldens-color}, 
where the electron density distribution along each PA is displayed and overlaid on the H$\alpha$ 
emission nebula. From this image it becomes also evident that higher densities occur in the regions 
of high intensities. Moreover, in the north-western region where the RVs are 
highest, the density rapidly decreases. The gaps, for example in the norther part of the PA at 
0$\degr$ and in the inner region of the PA at 295$\degr$ are caused by either an extremely faint, 
hence not measurable [S\,{\sc ii}] $\lambda$ 6731 line, or a contamination of that line by a strong 
cosmic ray, or a combination of both.

A few outliers appear with flux ratio values greater than the theoretically predicted one. We checked 
these flux measurements by eye, and they seem to be appropriate. Therefore, we believe that in these
extremely low-density regions our assumption of pure collisional excitation of the levels might not 
hold anymore.

\subsection{Analysis of the photometric light curve}
\label{sect:lightcurve}

\begin{figure*}
\begin{center}
\includegraphics[width=\hsize,angle=0]{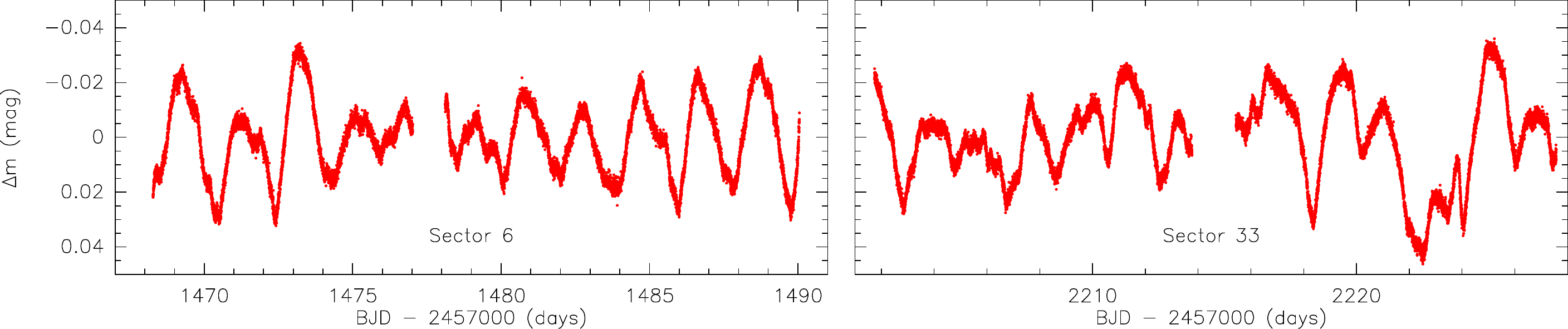}

\smallskip
\includegraphics[width=\hsize,angle=0]{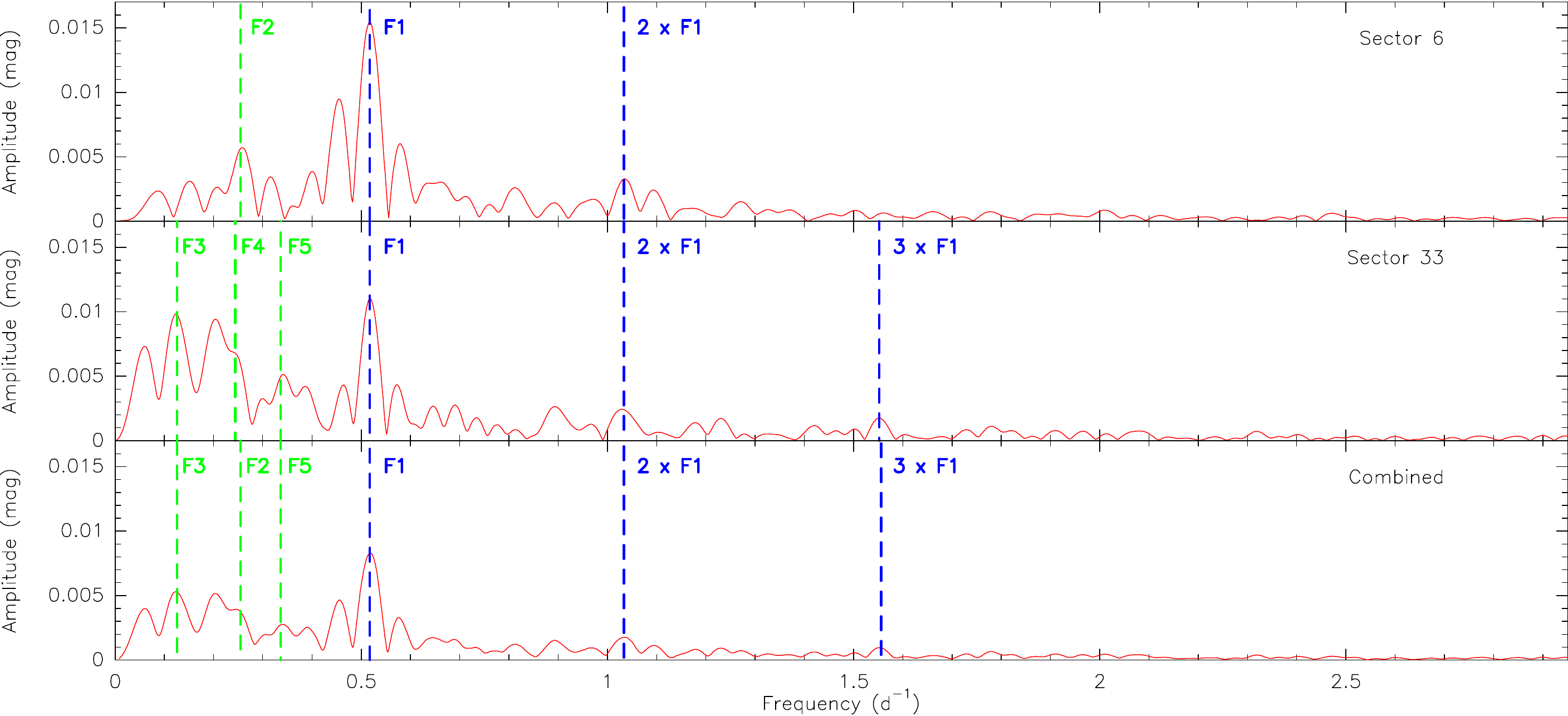}
\caption{TESS light curve of MWC~137 in the two sectors (top) and their first Fourier transform 
(bottom). Identified frequencies are indicated.}
\label{fig:TESS} 
\end{center}
\end{figure*}

The TESS time series of the central object MWC~137 taken in sector 6 and sector 33 are displayed in 
the top left and right panels of Figure~\ref{fig:TESS}, respectively. A slightly higher mean flux
value was noted for sector 33, corresponding to a difference of 0.022\,mag.
Considering the two year time gap between the observations, this brightening (if real and not just 
caused by very slight changes in the aperture and centering between sectors or due to contamination 
from nearby sources) might be related to a possible long-term variability of 332.4\,d. This period has 
been suggested by \citet{2018MNRAS.480..320M} based on their analysis of the ground-based light curve 
provided by the ASAS survey \citep{1997AcA....47..467P}, covering a total of about 2500\,d but with 
very sparse coverage. Such a long period cannot be recovered by the current set of TESS data. 
Therefore, we restrict our analysis to the light variability for each sector independently, 
considering their individual mean magnitudes.  

The two pieces of the light curve of MWC~137 display clear but complex variability. To search
for periodicities in these data sets, we perform a frequency analysis utilizing {\sc Period04} 
\citep{2005CoAst.146...53L} in a standard pre-whitening procedure. The amplitude spectra of the 
original data are shown in the lower parts of Figure~\ref{fig:TESS}. They have been computed
up to the Nyquist frequency ($\sim 360$\,d$^{-1}$ in both cases) and they display clear differences
in the signal strengths, especially in the low-frequency ($< 0.5$\,d$^{-1}$) domain. No significant 
peaks are detected beyond 2\,d$^{-1}$. We also include the amplitude spectrum of the combined data 
set (bottom panel).  

\begin{figure}
\begin{center}
\includegraphics[width=\hsize,angle=0]{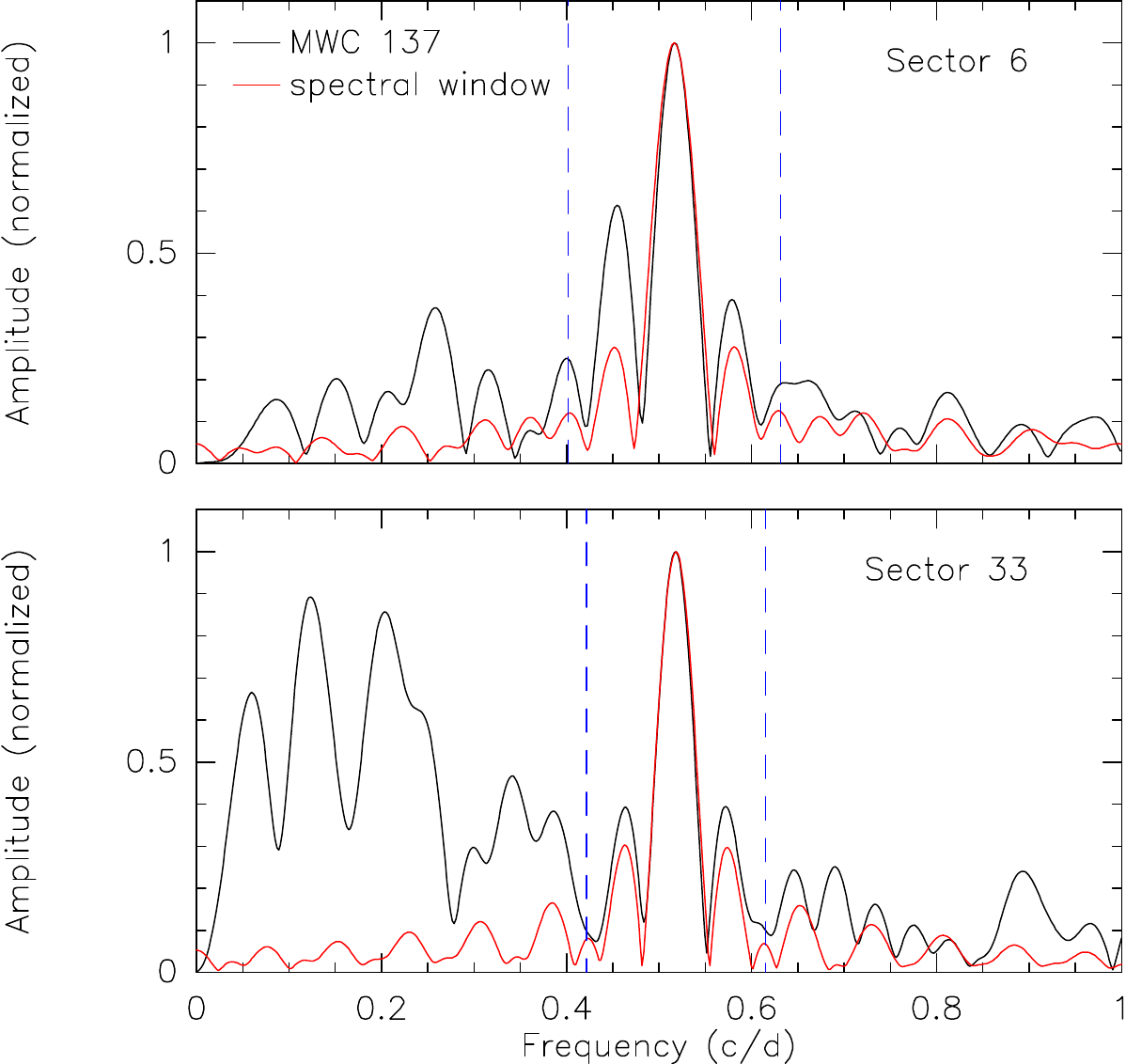}
\caption{Spectral window (red) for the two sectors centered on the frequency of the dominant peak in 
our original data (black). Also shown is the frequency resolution defined as 2.5 times the Rayleigh 
resolution (blue dashed lines).}
\label{fig:specwind} 
\end{center}
\end{figure}

The short observational baselines within the TESS sectors lead to several constraints for the 
identification of significant frequencies. First, it does not allow to measure signals with periods 
greater than $\sim 10$\,d, so that we ignore any detections in the low-frequency ($\leq 
0.1$\,d$^{-1}$) range. Second, while the Rayleigh resolution, defined as $1/T$ where $T$ is the time 
span of the observations, amounts to 0.046\,d$^{-1}$ and  0.039\,d$^{-1}$ for the data in sector 6 
and 33, respectively, high-amplitude side-lobes appear around dominant peaks in the amplitude spectra 
(see Figure~\ref{fig:specwind}). 
Therefore, we follow the conservative approach proposed by \citet{1978Ap&SS..56..285L} and discard 
frequencies with a separation of less than 2.5 times the Rayleigh resolution to previously identified 
frequencies with higher amplitude. Third, a higher threshold ($S/N \geq 5$) for the significance 
of identified peaks is recommended for an observational baseline of less than a few months of space 
photometry \citep{2015MNRAS.448L..16B}. And finally, a densely populated amplitude spectrum in the 
low-frequency domain requires a wider window size of 5\,d$^{-1}$ instead of the conventional 
1\,d$^{-1}$ for the computation of the underlying noise spectrum \citep{2020A&A...639A..81B}. 

Applying all these constraints during the frequency analysis, we end up with only a small number of 
frequencies that we consider as reliable. These frequencies are shown in the amplitude spectra of 
Figure~\ref{fig:TESS} and listed in Table\,\ref{tab:freq}. We clearly identify one dominant signal in 
both sectors with a period of 1.93\,d. The frequencies of its first two harmonics (2F1 and 3F1) are 
included in Figure~\ref{fig:TESS}. They are both seen in the amplitude spectrum of sector 33, whereas 
in sector 6 only the first harmonics is evident. Their signals fall slightly below our detection 
criterion ($4 < S/N < 5$) so that we include their frequencies only at the bottom of 
Table\,\ref{tab:freq}. The closeness of frequencies F2 in sector 6 and F4 in sector 33 suggests 
that they might be the same. If true, then this frequency could be another persistent one, while the 
other two frequencies (F3, F5) that show high amplitudes in sector 33, are not evident in sector 6 
and thus seem to be stochastic.

\section{Discussion}\label{sect:discussion}

\subsection{Large-scale nebula}

In a previous study \citet{2017AJ....154..186K} proposed a toy model for the nebula of MWC~137, 
consisting of two interwoven double-cones offset by $5\farcs 5$ in north-south direction.
The northern one was situated and sheared such that it fitted to the large-scale putative 
double-ring structure, whereas the southern one was oriented in a way that its symmetry axis was 
aligned with the jet axis. 

\begin{deluxetable}{lrr}[t]
\tablecaption{Significant frequencies identified in the TESS light curve. The harmonics of F1 are
listed at the bottom.
\label{tab:freq} }
\tablewidth{0pt}
\tablehead{
  & \colhead{Sector 6}    & \colhead{Sector 33}  \\
  & Frequency   & Frequency  \\ 
  & (d$^{-1}$)  & (d$^{-1}$)             
  }
\startdata
F1    &  0.51671400    & 0.51865874  \\
F2    &  0.25491224    &  \ldots \\
F3    &  \ldots        & 0.12579410  \\
F4    &  \ldots        & 0.24384702  \\
F5    &  \ldots        & 0.33674112  \\
\hline
2F1   &  1.03572450    & 1.03538218  \\
3F1   &   \ldots       & 1.55210563  \\
\enddata
\end{deluxetable}

Our new observations provide no complementary evidence that would support this model. Instead, the 
deeper image in the lines of [S\,{\sc ii}] imply that the large-scale nebula consists of 
helical or spiral-arm like structures rather than a double-ring (see 
Figure~\ref{fig:velocity_bins}). The RV measurements suggest that these helical 
structures are blue-shifted in the northern nebular regions and red-shifted in the southern ones, 
whereas the diffuse gas displays complex velocity variations. The lack of information about the 
velocity of the gas in the plane of the sky clearly hampers to draw conclusions about the real motion 
of the nebular structures in three dimensional space. But for the detection of the nebular expansion in 
the plane of the sky, our current baseline of 18.1\,yr is still too short.

\subsection{Stellar mass and age}

The significantly improved distance estimate towards MWC~137 based on Gaia, which agrees very well 
with the value derived by \citet{2016A&A...585A..81M}, allows us to finally settle the luminosity 
of the star and to better constrain its initial and current mass. The huge range of previously 
proposed distances resulted in luminosity estimates spreading from $\log (L/L_{\sun}) = 4.18$ 
\citep{2009A&A...497..117A} up to at least 5.37 \citep{1998MNRAS.298..185E}, and thus in masses 
ranging from 12 to at least 32\,$M_{\sun}$ when compared to stellar evolutionary 
tracks of rotating and non-rotating stars \citep[see][]{2015AJ....149...13M}. The most recent mass 
estimate has been performed by \citet{2016A&A...585A..81M}, who fit the location of MWC~137 in a 
color-magnitude diagram with GENEVA isochrones for the cluster and evolutionary tracks from 
\citet{2012A&A...537A.146E} based on which they claim best agreement for an age of $\sim 10$\,Myr and 
a mass of $\sim 15\,M_{\sun}$. At the same time, they estimate the luminosity of MWC~137 as $\log 
(L/L_{\sun}) \sim 6$, which clearly contradicts the low mass assigned to the object.

\begin{deluxetable*}{cccccccccccc}[t]
\tablecaption{Stellar parameters of MWC~137.
\label{tab:stelparam} }
\tablewidth{0pt}
\tablehead{
\colhead{$V$} & \colhead{$E(B-V)$}  & \colhead{$A_{V}$}  & \colhead{$d$} &  \colhead{$M_{V}$} & 
\colhead{$\log T_{\rm eff}$}  & \colhead{$\log L/L_{\sun}$} & \colhead{$M_{\rm in}$} & 
\colhead{$M$} & \colhead{$R$} &\colhead{$\Omega_{\rm in}/\Omega_{\rm{crit}}$} & \colhead{Age} \\
(mag) & (mag) &  (mag)  &  (kpc)  &  (mag)  & (K) &  &  ($M_{\sun}$) & ($M_{\sun}$) & ($R_{\sun}$) 
& & (Myr)   
}
\startdata
11.95$^{a}$ & 1.8$^{a}$ & 5.6$^{a}$ & 5.15$^{+0.79}_{-0.60}$ & $-7.2\pm 0.3$ & $4.40-4.49^{b}$ & 
$5.84\pm 0.13$ & $49\pm 11$ & $37^{+9}_{-5}$ & $25^{+20}_{-9}$ & $0.425\pm 0.025$ & $4.7\pm 0.8$ \\
\enddata
\tablereferences{$^{a}$Mehner et al. (2016); $^{b}$Hillenbrand et al. (1992); Esteban \& 
Fernandez (1998); Alonso-Albi et al. (2009).}
\end{deluxetable*} 

\begin{figure*}
\begin{center}
\includegraphics[width=\hsize,angle=0]{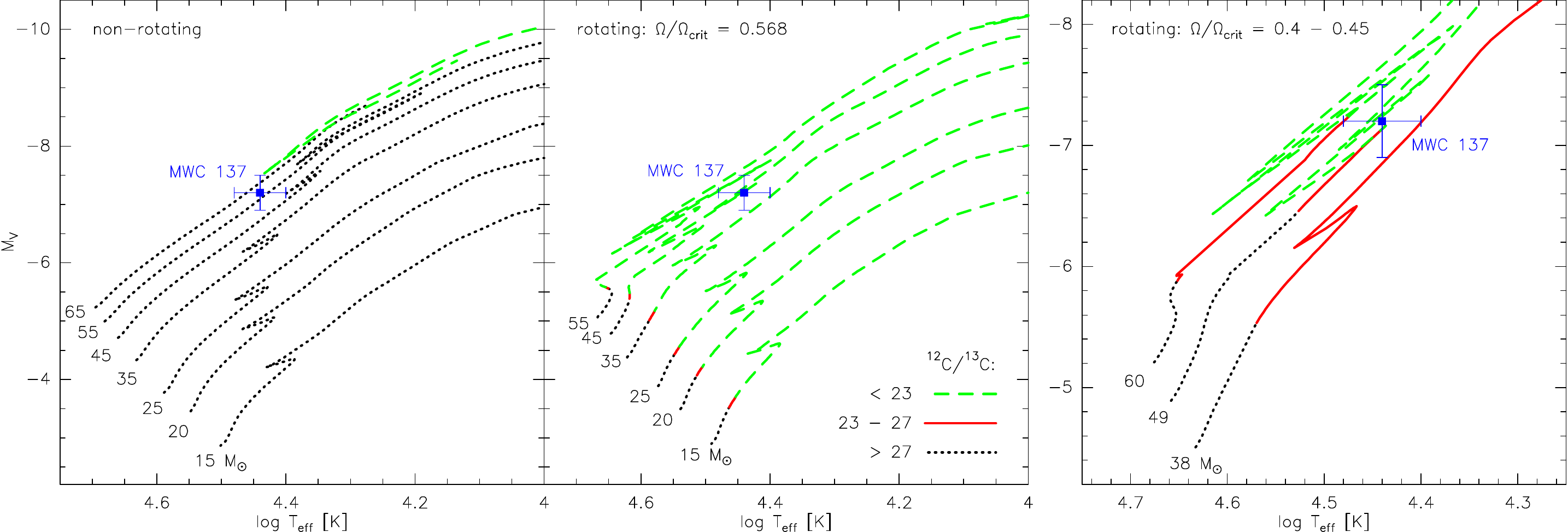}
\caption{Position of MWC~137 in the HR diagram. Evolutionary tracks for non-rotating (left) and 
rotating (middle) stellar models at solar metallicity are from \citet{2012A&A...537A.146E}. The color 
coding of the tracks refers to different ranges of the carbon isotope abundance ratio on the stellar 
surface with the solid red line covering the measurements for MWC~137. The right panel shows a zoomed 
version and evolutionary tracks for rotation speeds that predict the proper carbon isotope ratio at
the position of MWC~137.}
\label{fig:hrd} 
\end{center}
\end{figure*}

We believe that this mismatch between mass and luminosity arises from the fact that 
\citet{2016A&A...585A..81M} only matched the models with the absolute visual magnitude, $M_{V}$, 
but they neglected the effective temperature of the star. Temperature estimates in the literature all 
point towards a hot central object of $T_{\rm eff} = 25\,000 - 31\,000$\,K \citep[see, 
e.g.,][]{1992ApJ...397..613H, 1998MNRAS.298..185E, 2009A&A...497..117A}. Using the values of 
\citet{2016A&A...585A..81M} for the visual magnitude $V$ and the extinction $A_{V}$, along with the 
Gaia distance (Table\,\ref{tab:stelparam}), the absolute visual magnitude of MWC~137 is 
$M_{V} = -7.2\pm 0.3$\,mag, in excellent agreement with the value of \citet{2016A&A...585A..81M}.

In Figure~\ref{fig:hrd} we display evolutionary tracks for non-rotating (left panel) and rotating 
stellar models with $\Omega_{\rm in}/\Omega_{\rm crit} = 0.568$ (respectively $v/v_{\rm crit} = 0.4$, 
middle panel) from \citet{2012A&A...537A.146E} for solar metallicity. If MWC~137 had an initial mass 
of $15\,M_{\sun}$ as proposed by \citet{2016A&A...585A..81M}, a value of $M_{V} = -7.2$\,mag is 
achieved only when the star has reached an effective temperature $\log T_{\rm eff} \le 4$ according 
to the models of both rotating and non-rotating stars. Considering the proper effective temperature 
range for MWC~137 of $\log T_{\rm eff} = 4.40 - 4.49$, the real initial mass is considerably higher 
than $15\,M_{\sun}$, spreading from about 35 to 65\,$M_{\sun}$.

To further narrow down the mass range, we included in Figure~\ref{fig:hrd} information about the 
surface abundance ratio of $^{13}$C/$^{12}$C. This ratio is very sensitive to stellar rotation and 
may help to make inferences about the initial rotation speed of MWC~137. While we cannot measure the
carbon isotope abundance ratio directly on the stellar surface, \citet{2009A&A...494..253K} has shown 
that the enrichment of the close-by circumstellar environment in the isotopic molecule $^{13}$CO 
provides a reasonable tracer of the stellar surface enrichment in $^{13}$C for stars with strong 
winds or mass ejections. The circumstellar matter of MWC~137 indicates a carbon isotope abundance 
ratio of $^{13}$C/$^{12}$C = $25\pm2$ \citep{2015AJ....149...13M, 2013A&A...558A..17O}. While the 
non-rotating tracks do not display significant surface enrichment at early stellar 
evolutionary phases, the models with rotation predict that the amount observed in MWC~137 is reached 
already during the main sequence and significantly lower around the position of MWC~137 in the HR 
diagram. Therefore, to achieve the observed value, the initial rotation rate of MWC~137 must be 
clearly non-zero but also lower than $\Omega_{\rm in}/\Omega_{\rm crit} = 0.568$. 

We interpolate the pre-calculated evolution tracks of \citet{2012A&A...537A.146E} using their tool 
{\sc SYCLIST\footnote{https://www.unige.ch/sciences/astro/evolution/en/database/ syclist/}} to 
obtain stellar evolution tracks for a dense mass grid and covering rotation rates between
$\Omega_{\rm in}/\Omega_{\rm crit} = 0$ and 0.568. The evolution of the carbon isotope ratio is 
followed along these tracks to select the best fitting models for the parameters of MWC~137. We find, 
that the most likely initial mass of MWC~137 is $49\pm 11\,M_{\sun}$ with an initial rotation rate of 
$\Omega_{\rm in}/\Omega_{\rm crit} = 0.40 - 0.43$. Slightly higher rotation rate values (up to 
0.45) are still possible for models at the upper mass limit. Representative tracks are shown in the 
right panel of Figure~\ref{fig:hrd}. From these tracks we deduce a current stellar mass of 
$37^{+9}_{-5}\,M_{\sun}$ and a luminosity of $\log (L/L_{\sun}) = 5.84\pm 0.13$ for MWC~137. The 
latter is in good agreement with the luminosity of $\log (L/L_{\sun}) \sim 6$ derived by 
\citet{2016A&A...585A..81M}. The final set of stellar parameters of MWC~137 is presented in 
Table\,\ref{tab:stelparam}. The table also includes the plausible age of MWC~137, which is found to 
range from 3.9 to 5.5\,Myr, in excellent agreement with the estimated ages of the other stars in that 
cluster \citep[see][]{2016A&A...585A..81M}.

\subsection{Photometric variability}

The TESS time series, taken in sectors 6 and 33 with a separation of two years, contains at least 
one stable period of 1.93\,d (see Sect.\,\ref{sect:lightcurve}). Its putative decrease in amplitude 
between the two observing runs (see Figure~\ref{fig:TESS}) might not be real but caused by the 
simultaneous weakening or disappearance of possible, unresolved stochastic signals in the vicinity of 
the dominant frequency. Considering the stellar parameters of MWC~137 (see Table~\ref{tab:stelparam}), 
it is obvious that a period of 1.93\,d is way too short to be interpreted as the rotation velocity of 
the star. The eligible stellar models predict an equatorial rotation velocity of 
$< 10$\,km\,s$^{-1}$ for the current evolutionary state of MWC~137, resulting in a minimum stellar 
rotation period of $\sim 100$\,d. It is also unlikely that the 1.93\,d period could be caused by the 
orbital motion of a hidden companion. Considering the parameter range for MWC~137, the corresponding 
orbital radius of the companion would lie inside the stellar radius. Therefore, a different alternative 
for the appearance of the 1.93\,d period seems to be more likely, and this is stellar pulsation.

The other, stochastic or random signals seen in the amplitude spectrum of MWC~137 might be related to 
internal gravity waves. The appearance of such waves has been suggested for early-type stars, and 
they seem to be suitable to generate stochastic photometric variability \citep{2013ApJ...772...21R}. 
The clearly variable amplitude and evanescence of these signals further support such an 
interpretation. 

But other or additional mechanisms might be at work and are worth being mentioned. 
As a massive, early-type star MWC~137 has a line-driven wind. Recent investigations have shown that 
perturbations in the velocity field at the base of the wind, for example by stochastic waves, can 
cause wind instabilities \citep{2021A&A...648A..79K}. These instabilities can generate variability 
of the mass flux and hence wind inhomogeneities that might lead to measurable (quasi-periodic or 
stochastic) changes in photometry. A time variable wind of MWC~137 is evident from moderate changes 
in its hydrogen Balmer line emission at optical wavelengths \citep{2003A&A...408..257Z, 
2017AJ....154..186K} as well as from variable Pfund line emission in the near-infrared 
\citep{2015AJ....149...13M}. Furthermore, the central star is surrounded by a compact region of 
molecular and neutral gas, possibly arranged in a series of rings \citep{2018MNRAS.480..320M}, and 
the absence of photometric lines in the spectra implies that the star is embedded in a dense 
envelope. A compact disk has also been proposed to be the source of the emanating jet and blobs 
discovered by \citet{2016A&A...585A..81M}, and dynamical dusty blobs and arc-like structures have 
been found in the close vicinity of MWC~137 as well \citep{2017AJ....154..186K}. Therefore, it is 
clear that any time-variable structures around the central star can easily imprint variabilities on 
the observed photometric signals.

\section{Conclusions}\label{sect:conclusions}

In this work, we present new optical observations of the large-scale nebula connected to the B[e] 
supergiant star MWC~137. We find no evidence for an expansion of the nebular structures in the plane 
of the sky based on H$\alpha$ images spanning over more than 18\,yr. This means that the tangential 
velocity must be smaller than $\sim 53$\,km\,s$^{-1}$ for the distance of 5.15\,kpc to MWC~137. The 
RVs measured from the [N\,{\sc ii}] and [S\,{\sc ii}] emission lines reveal a complex pattern across 
the nebula. But in general, the data confirm the previous findings that the northern nebular regions 
are predominantly approaching whereas the southern ones are mostly receding. We also find that the 
velocity dispersion is highest in regions of highest RV. Moreover, the RV measurements reveal 
systematically higher values for the [N\,{\sc ii}] lines in the blue-shifted north and north-western 
regions as well as in the red-shifted south-eastern area in which the RVs are highest. These regions 
coincide with the regions of lowest electron densities. The clear correlation of higher RV with higher 
ionization potential of the element suggests a velocity stratification along the line of sight, meaning 
that we observe a decrease in RV from inside out. Whether this decrease is consistent with ballistic 
expansion needs to be checked once tangential velocities have been measured. Proper fully 3D kinematics 
will also help to understand the formation scenario of the diverse observed spiral-arm like structures.

In addition to the spectroscopic data, we also investigate the photometric light curve provided by the 
TESS space mission. Our analysis reveals a few low-frequency stochastic or random signals and one 
stable 1.93\,d period over the observing baseline of 2\,yr. This period can neither be related to the 
rotation velocity of the star, which is minimum 100\,d considering our derived mass of $\sim 
37$\,M$_{\odot}$, luminosity of $\log L/L_{\odot} \sim 5.8$, and age of $3.9-5.5$\,Myr of MWC~137, 
nor can it be interpreted as orbital motion of a hidden companion. Therefore, stellar pulsation might 
be a plausible alternative. The real nature of both the 1.93\,d and the low-frequency stochastic 
variability certainly requires further investigation, including computations of stellar pulsation 
models for the parameter space of MWC~137. However, this is beyond the scope of the present work.

\acknowledgments

This research made use of the NASA Astrophysics Data System (ADS), of the 
SIMBAD database, operated at CDS, Strasbourg, France, and of data from the European Space Agency (ESA) 
mission {\it Gaia} (\url{https://www.cosmos.esa.int/gaia}), processed by the {\it Gaia}
Data Processing and Analysis Consortium (DPAC,
\url{https://www.cosmos.esa.int/web/gaia/dpac/consortium}). Funding for the DPAC
has been provided by national institutions, in particular the institutions
participating in the {\it Gaia} Multilateral Agreement. 
Parts of the data presented here were obtained with ALFOSC, which is provided by the Instituto de 
Astrofisica de Andalucia (IAA) under a joint agreement with the University of Copenhagen and NOTSA, 
and with the 6-m telescope of the Special Astrophysical Observatory, operated with the financial 
support of the Ministry of Education and Science of the Russian Federation (agreement No. 
14.619.21.0004, project ID RFMEFI61914X0004).
This paper includes also data collected with the TESS mission, obtained from the MAST data archive at 
the Space Telescope Science Institute (STScI). Funding for the TESS mission is provided by the NASA 
Explorer Program. STScI is operated by the Association of Universities for Research in Astronomy, 
Inc., under NASA contract NAS 5-26555.

We wish to thank Miguel Santander-Garc\'{i}a for advice on using the magnification method, and Anthony 
Marston for providing us with his H$\alpha$ image.  
MK, JPSA, and DHN acknowledge financial support from the Czech Science Foundation (GA\,\v{C}R 
20-00150S). The Astronomical Institute Ond\v{r}ejov is supported by the project RVO:67985815. 
LSC acknowledges financial support from CONICET (PIP 0177) and
from the Agencia Nacional de Promoci\'{o}n Cient\'{i}fica y Tecnol\'{o}gica de Argentina
(Pr\'{e}stamo BID PICT 2016-1971)
DJ acknowledges support from the Erasmus+ programme of the European Union under grant number 
2020-1-CZ01-KA203-078200. This project has received funding from the European Union's Framework 
Programme for Research and Innovation Horizon 2020 (2014-2020) under the Marie Sk\l{}odowska-Curie 
Grant Agreement No. 823734.

\facility{NOT}
\facility{BAT}
\facility{Danish 1.54m Telescope}

\bibliographystyle{aasjournal}
\bibliography{ms}

\begin{thebibliography}{}
\expandafter\ifx\csname natexlab\endcsname\relax\def\natexlab#1{#1}\fi
\providecommand{\url}[1]{\href{#1}{#1}}
\providecommand{\dodoi}[1]{doi:~\href{http://doi.org/#1}{\nolinkurl{#1}}}
\providecommand{\doeprint}[1]{\href{http://ascl.net/#1}{\nolinkurl{http://ascl.net/#1}}}
\providecommand{\doarXiv}[1]{\href{https://arxiv.org/abs/#1}{\nolinkurl{https://arxiv.org/abs/#1}}}

\bibitem[{{Ababakr} {et~al.}(2017){Ababakr}, {Oudmaijer}, \&
  {Vink}}]{2017MNRAS.472..854A}
{Ababakr}, K.~M., {Oudmaijer}, R.~D., \& {Vink}, J.~S. 2017, \mnras, 472, 854,
  \dodoi{10.1093/mnras/stx1891}

\bibitem[{{Afanasiev} \& {Moiseev}(2011)}]{2011BaltA..20..363A}
{Afanasiev}, V.~L., \& {Moiseev}, A.~V. 2011, Baltic Astronomy, 20, 363,
  \dodoi{10.1515/astro-2017-0305}

\bibitem[{{Allen} \& {Glass}(1976)}]{1976ApJ...210..666A}
{Allen}, D.~A., \& {Glass}, I.~S. 1976, \apj, 210, 666, \dodoi{10.1086/154872}

\bibitem[{{Alonso-Albi} {et~al.}(2009){Alonso-Albi}, {Fuente}, {Bachiller},
  {Neri}, {Planesas}, {Testi}, {Bern{\'e}}, \& {Joblin}}]{2009A&A...497..117A}
{Alonso-Albi}, T., {Fuente}, A., {Bachiller}, R., {et~al.} 2009, \aap, 497,
  117, \dodoi{10.1051/0004-6361/200810401}

\bibitem[{{Arellano-C{\'o}rdova} {et~al.}(2020){Arellano-C{\'o}rdova},
  {Esteban}, {Garc{\'\i}a-Rojas}, \&
  {M{\'e}ndez-Delgado}}]{2020MNRAS.496.1051A}
{Arellano-C{\'o}rdova}, K.~Z., {Esteban}, C., {Garc{\'\i}a-Rojas}, J., \&
  {M{\'e}ndez-Delgado}, J.~E. 2020, \mnras, 496, 1051,
  \dodoi{10.1093/mnras/staa1523}

\bibitem[{{Aret} {et~al.}(2012){Aret}, {Kraus}, {Muratore}, \& {Borges
  Fernandes}}]{2012MNRAS.423..284A}
{Aret}, A., {Kraus}, M., {Muratore}, M.~F., \& {Borges Fernandes}, M. 2012,
  \mnras, 423, 284, \dodoi{10.1111/j.1365-2966.2012.20871.x}

\bibitem[{{Arun} {et~al.}(2019){Arun}, {Mathew}, {Manoj}, {Ujjwal}, {Kartha},
  {Viswanath}, {Narang}, \& {Paul}}]{2019AJ....157..159A}
{Arun}, R., {Mathew}, B., {Manoj}, P., {et~al.} 2019, \aj, 157, 159,
  \dodoi{10.3847/1538-3881/ab0ca1}

\bibitem[{{Baran} {et~al.}(2015){Baran}, {Koen}, \&
  {Pokrzywka}}]{2015MNRAS.448L..16B}
{Baran}, A.~S., {Koen}, C., \& {Pokrzywka}, B. 2015, \mnras, 448, L16,
  \dodoi{10.1093/mnrasl/slu194}

\bibitem[{{Berrilli} {et~al.}(1992){Berrilli}, {Corciulo}, {Ingrosso},
  {Lorenzetti}, {Nisini}, \& {Strafella}}]{1992ApJ...398..254B}
{Berrilli}, F., {Corciulo}, G., {Ingrosso}, G., {et~al.} 1992, \apj, 398, 254,
  \dodoi{10.1086/171853}

\bibitem[{{Bonanos} {et~al.}(2009){Bonanos}, {Massa}, {Sewilo}, {Lennon},
  {Panagia}, {Smith}, {Meixner}, {Babler}, {Bracker}, {Meade}, {Gordon},
  {Hora}, {Indebetouw}, \& {Whitney}}]{2009AJ....138.1003B}
{Bonanos}, A.~Z., {Massa}, D.~L., {Sewilo}, M., {et~al.} 2009, \aj, 138, 1003,
  \dodoi{10.1088/0004-6256/138/4/1003}

\bibitem[{{Bonanos} {et~al.}(2010){Bonanos}, {Lennon}, {K{\"o}hlinger}, {van
  Loon}, {Massa}, {Sewilo}, {Evans}, {Panagia}, {Babler}, {Block}, {Bracker},
  {Engelbracht}, {Gordon}, {Hora}, {Indebetouw}, {Meade}, {Meixner}, {Misselt},
  {Robitaille}, {Shiao}, \& {Whitney}}]{2010AJ....140..416B}
{Bonanos}, A.~Z., {Lennon}, D.~J., {K{\"o}hlinger}, F., {et~al.} 2010, \aj,
  140, 416, \dodoi{10.1088/0004-6256/140/2/416}

\bibitem[{{Brandner} {et~al.}(1997){Brandner}, {Grebel}, {Chu}, \&
  {Weis}}]{1997ApJ...475L..45B}
{Brandner}, W., {Grebel}, E.~K., {Chu}, Y.-H., \& {Weis}, K. 1997, \apjl, 475,
  L45, \dodoi{10.1086/310460}

\bibitem[{{Burssens} {et~al.}(2020){Burssens}, {Sim{\'o}n-D{\'\i}az}, {Bowman},
  {Holgado}, {Michielsen}, {de Burgos}, {Castro}, {Barb{\'a}}, \&
  {Aerts}}]{2020A&A...639A..81B}
{Burssens}, S., {Sim{\'o}n-D{\'\i}az}, S., {Bowman}, D.~M., {et~al.} 2020,
  \aap, 639, A81, \dodoi{10.1051/0004-6361/202037700}

\bibitem[{{Caicedo-Ortiz} \& {Fernandez}(2019)}]{2019Ap.....62...57C}
{Caicedo-Ortiz}, H.~E., \& {Fernandez}, H.~O.~C. 2019, Astrophysics, 62, 57,
  \dodoi{10.1007/s10511-019-09564-9}

\bibitem[{{Cidale} {et~al.}(2012){Cidale}, {Borges Fernandes}, {Andruchow},
  {Arias}, {Kraus}, {Chesneau}, {Kanaan}, {Cur{\'e}}, {de Wit}, \&
  {Muratore}}]{2012A&A...548A..72C}
{Cidale}, L.~S., {Borges Fernandes}, M., {Andruchow}, I., {et~al.} 2012, \aap,
  548, A72, \dodoi{10.1051/0004-6361/201220120}

\bibitem[{{Cochetti} {et~al.}(2020){Cochetti}, {Kraus}, {Arias}, {Cidale},
  {Eenm{\"a}e}, {Liimets}, {Torres}, \& {Djupvik}}]{2020AJ....160..166C}
{Cochetti}, Y.~R., {Kraus}, M., {Arias}, M.~L., {et~al.} 2020, \aj, 160, 166,
  \dodoi{10.3847/1538-3881/abae62}

\bibitem[{{Ekstr{\"o}m} {et~al.}(2012){Ekstr{\"o}m}, {Georgy}, {Eggenberger},
  {Meynet}, {Mowlavi}, {Wyttenbach}, {Granada}, {Decressin}, {Hirschi},
  {Frischknecht}, {Charbonnel}, \& {Maeder}}]{2012A&A...537A.146E}
{Ekstr{\"o}m}, S., {Georgy}, C., {Eggenberger}, P., {et~al.} 2012, \aap, 537,
  A146, \dodoi{10.1051/0004-6361/201117751}

\bibitem[{{Esteban} \& {Fernandez}(1998)}]{1998MNRAS.298..185E}
{Esteban}, C., \& {Fernandez}, M. 1998, \mnras, 298, 185,
  \dodoi{10.1046/j.1365-8711.1998.01610.x}

\bibitem[{{Esteban} \& {Peimbert}(1999)}]{1999A&A...349..276E}
{Esteban}, C., \& {Peimbert}, M. 1999, \aap, 349, 276

\bibitem[{{Fich} \& {Blitz}(1984)}]{1984ApJ...279..125F}
{Fich}, M., \& {Blitz}, L. 1984, \apj, 279, 125, \dodoi{10.1086/161872}

\bibitem[{{Fuente} {et~al.}(2003){Fuente}, {Rodr{\'\i}guez-Franco}, {Testi},
  {Natta}, {Bachiller}, \& {Neri}}]{2003ApJ...598L..39F}
{Fuente}, A., {Rodr{\'\i}guez-Franco}, A., {Testi}, L., {et~al.} 2003, \apjl,
  598, L39, \dodoi{10.1086/380296}

\bibitem[{{Gaia Collaboration} {et~al.}(2016){Gaia Collaboration}, {Prusti},
  {de Bruijne}, {Brown}, {Vallenari}, {Babusiaux}, {Bailer-Jones}, {Bastian},
  {Biermann}, {Evans}, \& et~al.}]{2016A&A...595A...1G}
{Gaia Collaboration}, {Prusti}, T., {de Bruijne}, J.~H.~J., {et~al.} 2016,
  \aap, 595, A1, \dodoi{10.1051/0004-6361/201629272}

\bibitem[{{Gaia Collaboration} {et~al.}(2021){Gaia Collaboration}, {Brown},
  {Vallenari}, {Prusti}, {de Bruijne}, {Babusiaux}, {Biermann}, {Creevey},
  {Evans}, {Eyer}, \& et~al.}]{2021A&A...649A...1G}
{Gaia Collaboration}, {Brown}, A.~G.~A., {Vallenari}, A., {et~al.} 2021, \aap,
  649, A1, \dodoi{10.1051/0004-6361/202039657}

\bibitem[{{Hillenbrand} {et~al.}(1992){Hillenbrand}, {Strom}, {Vrba}, \&
  {Keene}}]{1992ApJ...397..613H}
{Hillenbrand}, L.~A., {Strom}, S.~E., {Vrba}, F.~J., \& {Keene}, J. 1992, \apj,
  397, 613, \dodoi{10.1086/171819}

\bibitem[{{Kraus}(2009)}]{2009A&A...494..253K}
{Kraus}, M. 2009, \aap, 494, 253, \dodoi{10.1051/0004-6361:200811020}

\bibitem[{{Kraus}(2019)}]{2019Galax...7...83K}
---. 2019, Galaxies, 7, 83, \dodoi{10.3390/galaxies7040083}

\bibitem[{{Kraus} {et~al.}(2020){Kraus}, {Arias}, {Cidale}, \&
  {Torres}}]{2020MNRAS.493.4308K}
{Kraus}, M., {Arias}, M.~L., {Cidale}, L.~S., \& {Torres}, A.~F. 2020, \mnras,
  493, 4308, \dodoi{10.1093/mnras/staa519}

\bibitem[{{Kraus} {et~al.}(2007){Kraus}, {Borges Fernandes}, \& {de
  Ara{\'u}jo}}]{2007A&A...463..627K}
{Kraus}, M., {Borges Fernandes}, M., \& {de Ara{\'u}jo}, F.~X. 2007, \aap, 463,
  627, \dodoi{10.1051/0004-6361:20066325}

\bibitem[{{Kraus} {et~al.}(2010){Kraus}, {Borges Fernandes}, \& {de
  Ara{\'u}jo}}]{2010A&A...517A..30K}
---. 2010, \aap, 517, A30, \dodoi{10.1051/0004-6361/200913964}

\bibitem[{{Kraus} {et~al.}(2014){Kraus}, {Cidale}, {Arias}, {Oksala}, \&
  {Borges Fernandes}}]{2014ApJ...780L..10K}
{Kraus}, M., {Cidale}, L.~S., {Arias}, M.~L., {Oksala}, M.~E., \& {Borges
  Fernandes}, M. 2014, \apjl, 780, L10, \dodoi{10.1088/2041-8205/780/1/L10}

\bibitem[{{Kraus} {et~al.}(2015){Kraus}, {Oksala}, {Cidale}, {Arias}, {Torres},
  \& {Borges Fernandes}}]{2015ApJ...800L..20K}
{Kraus}, M., {Oksala}, M.~E., {Cidale}, L.~S., {et~al.} 2015, \apjl, 800, L20,
  \dodoi{10.1088/2041-8205/800/2/L20}

\bibitem[{{Kraus} {et~al.}(2016){Kraus}, {Cidale}, {Arias}, {Maravelias},
  {Nickeler}, {Torres}, {Borges Fernandes}, {Aret}, {Cur{\'e}},
  {Vallverd{\'u}}, \& {Barb{\'a}}}]{2016A&A...593A.112K}
{Kraus}, M., {Cidale}, L.~S., {Arias}, M.~L., {et~al.} 2016, \aap, 593, A112,
  \dodoi{10.1051/0004-6361/201628493}

\bibitem[{{Kraus} {et~al.}(2017){Kraus}, {Liimets}, {Cappa}, {Cidale},
  {Nickeler}, {Duronea}, {Arias}, {Gunawan}, {Oksala}, {Fernandes},
  {Maravelias}, {Cur{\'e}}, \& {Santander-Garc{\'\i}a}}]{2017AJ....154..186K}
{Kraus}, M., {Liimets}, T., {Cappa}, C.~E., {et~al.} 2017, \aj, 154, 186,
  \dodoi{10.3847/1538-3881/aa8df6}

\bibitem[{{Krti{\v{c}}ka} \& {Feldmeier}(2021)}]{2021A&A...648A..79K}
{Krti{\v{c}}ka}, J., \& {Feldmeier}, A. 2021, \aap, 648, A79,
  \dodoi{10.1051/0004-6361/202040148}

\bibitem[{{Lamers} {et~al.}(1998){Lamers}, {Zickgraf}, {de Winter}, {Houziaux},
  \& {Zorec}}]{1998A&A...340..117L}
{Lamers}, H. J.~G.~L.~M., {Zickgraf}, F.-J., {de Winter}, D., {Houziaux}, L.,
  \& {Zorec}, J. 1998, \aap, 340, 117

\bibitem[{{Lenz} \& {Breger}(2005)}]{2005CoAst.146...53L}
{Lenz}, P., \& {Breger}, M. 2005, Communications in Asteroseismology, 146, 53,
  \dodoi{10.1553/cia146s53}

\bibitem[{{Liermann} {et~al.}(2010){Liermann}, {Kraus}, {Schnurr}, \&
  {Fernandes}}]{2010MNRAS.408L...6L}
{Liermann}, A., {Kraus}, M., {Schnurr}, O., \& {Fernandes}, M.~B. 2010, \mnras,
  408, L6, \dodoi{10.1111/j.1745-3933.2010.00915.x}

\bibitem[{{Liermann} {et~al.}(2014){Liermann}, {Schnurr}, {Kraus}, {Kreplin},
  {Arias}, \& {Cidale}}]{2014MNRAS.443..947L}
{Liermann}, A., {Schnurr}, O., {Kraus}, M., {et~al.} 2014, \mnras, 443, 947,
  \dodoi{10.1093/mnras/stu1174}

\bibitem[{{Liimets} {et~al.}(2018){Liimets}, {Corradi}, {Jones}, {Verro},
  {Santander-Garc{\'\i}a}, {Kolka}, {Sidonio}, {Kankare}, {Kankare}, {Pursimo},
  \& {Wilson}}]{2018A&A...612A.118L}
{Liimets}, T., {Corradi}, R.~L.~M., {Jones}, D., {et~al.} 2018, \aap, 612,
  A118, \dodoi{10.1051/0004-6361/201732073}

\bibitem[{{Loumos} \& {Deeming}(1978)}]{1978Ap&SS..56..285L}
{Loumos}, G.~L., \& {Deeming}, T.~J. 1978, \apss, 56, 285,
  \dodoi{10.1007/BF01879560}

\bibitem[{{Luri} {et~al.}(2018){Luri}, {Brown}, {Sarro}, {Arenou},
  {Bailer-Jones}, {Castro-Ginard}, {de Bruijne}, {Prusti}, {Babusiaux}, \&
  {Delgado}}]{2018A&A...616A...9L}
{Luri}, X., {Brown}, A.~G.~A., {Sarro}, L.~M., {et~al.} 2018, \aap, 616, A9,
  \dodoi{10.1051/0004-6361/201832964}

\bibitem[{{Maravelias} {et~al.}(2018){Maravelias}, {Kraus}, {Cidale}, {Borges
  Fernandes}, {Arias}, {Cur{\'e}}, \& {Vasilopoulos}}]{2018MNRAS.480..320M}
{Maravelias}, G., {Kraus}, M., {Cidale}, L.~S., {et~al.} 2018, \mnras, 480,
  320, \dodoi{10.1093/mnras/sty1747}

\bibitem[{{Marston} \& {McCollum}(2008)}]{2008A&A...477..193M}
{Marston}, A.~P., \& {McCollum}, B. 2008, \aap, 477, 193,
  \dodoi{10.1051/0004-6361:20066086}

\bibitem[{{McGregor} {et~al.}(1988){McGregor}, {Hyland}, \&
  {Hillier}}]{1988ApJ...334..639M}
{McGregor}, P.~J., {Hyland}, A.~R., \& {Hillier}, D.~J. 1988, \apj, 334, 639,
  \dodoi{10.1086/166867}

\bibitem[{{Mehner} {et~al.}(2016){Mehner}, {de Wit}, {Groh}, {Oudmaijer},
  {Baade}, {Rivinius}, {Selman}, {Boffin}, \& {Martayan}}]{2016A&A...585A..81M}
{Mehner}, A., {de Wit}, W.~J., {Groh}, J.~H., {et~al.} 2016, \aap, 585, A81,
  \dodoi{10.1051/0004-6361/201527180}

\bibitem[{{Moiseev}(2002)}]{2002BSAO...54...74M}
{Moiseev}, A.~V. 2002, Bulletin of the Special Astrophysics Observatory, 54,
  74.
\newblock \doarXiv{astro-ph/0211104}

\bibitem[{{Moiseev}(2015)}]{2015AstBu..70..494M}
---. 2015, Astrophysical Bulletin, 70, 494, \dodoi{10.1134/S1990341315040112}

\bibitem[{{Moiseev} \& {Egorov}(2008)}]{2008AstBu..63..181M}
{Moiseev}, A.~V., \& {Egorov}, O.~V. 2008, Astrophysical Bulletin, 63, 181,
  \dodoi{10.1134/S1990341308020089}

\bibitem[{{Morris} {et~al.}(1996){Morris}, {Eenens}, {Hanson}, {Conti}, \&
  {Blum}}]{1996ApJ...470..597M}
{Morris}, P.~W., {Eenens}, P.~R.~J., {Hanson}, M.~M., {Conti}, P.~S., \&
  {Blum}, R.~D. 1996, \apj, 470, 597, \dodoi{10.1086/177892}

\bibitem[{{Muratore} {et~al.}(2015){Muratore}, {Kraus}, {Oksala}, {Arias},
  {Cidale}, {Borges Fernandes}, \& {Liermann}}]{2015AJ....149...13M}
{Muratore}, M.~F., {Kraus}, M., {Oksala}, M.~E., {et~al.} 2015, \aj, 149, 13,
  \dodoi{10.1088/0004-6256/149/1/13}

\bibitem[{{Oksala} {et~al.}(2012){Oksala}, {Kraus}, {Arias}, {Borges
  Fernandes}, {Cidale}, {Muratore}, \& {Cur{\'e}}}]{2012MNRAS.426L..56O}
{Oksala}, M.~E., {Kraus}, M., {Arias}, M.~L., {et~al.} 2012, \mnras, 426, L56,
  \dodoi{10.1111/j.1745-3933.2012.01323.x}

\bibitem[{{Oksala} {et~al.}(2013){Oksala}, {Kraus}, {Cidale}, {Muratore}, \&
  {Borges Fernandes}}]{2013A&A...558A..17O}
{Oksala}, M.~E., {Kraus}, M., {Cidale}, L.~S., {Muratore}, M.~F., \& {Borges
  Fernandes}, M. 2013, \aap, 558, A17, \dodoi{10.1051/0004-6361/201321568}

\bibitem[{{Oudmaijer} \& {Drew}(1999)}]{1999MNRAS.305..166O}
{Oudmaijer}, R.~D., \& {Drew}, J.~E. 1999, \mnras, 305, 166,
  \dodoi{10.1046/j.1365-8711.1999.02383.x}

\bibitem[{{Pojmanski}(1997)}]{1997AcA....47..467P}
{Pojmanski}, G. 1997, \actaa, 47, 467.
\newblock \doarXiv{astro-ph/9712146}

\bibitem[{{Reed} {et~al.}(1999){Reed}, {Balick}, {Hajian}, {Klayton},
  {Giovanardi}, {Casertano}, {Panagia}, \& {Terzian}}]{1999AJ....118.2430R}
{Reed}, D.~S., {Balick}, B., {Hajian}, A.~R., {et~al.} 1999, \aj, 118, 2430,
  \dodoi{10.1086/301091}

\bibitem[{{Rogers} {et~al.}(2013){Rogers}, {Lin}, {McElwaine}, \&
  {Lau}}]{2013ApJ...772...21R}
{Rogers}, T.~M., {Lin}, D.~N.~C., {McElwaine}, J.~N., \& {Lau}, H.~H.~B. 2013,
  \apj, 772, 21, \dodoi{10.1088/0004-637X/772/1/21}

\bibitem[{{Sandell} {et~al.}(2011){Sandell}, {Weintraub}, \&
  {Hamidouche}}]{2011ApJ...727...26S}
{Sandell}, G., {Weintraub}, D.~A., \& {Hamidouche}, M. 2011, \apj, 727, 26,
  \dodoi{10.1088/0004-637X/727/1/26}

\bibitem[{{Smith} {et~al.}(2007){Smith}, {Bally}, \&
  {Walawender}}]{2007AJ....134..846S}
{Smith}, N., {Bally}, J., \& {Walawender}, J. 2007, \aj, 134, 846,
  \dodoi{10.1086/518563}

\bibitem[{{Tayal} \& {Zatsarinny}(2010)}]{2010ApJS..188...32T}
{Tayal}, S.~S., \& {Zatsarinny}, O. 2010, \apjs, 188, 32,
  \dodoi{10.1088/0067-0049/188/1/32}

\bibitem[{{The} {et~al.}(1994){The}, {de Winter}, \&
  {Perez}}]{1994A&AS..104..315T}
{The}, P.~S., {de Winter}, D., \& {Perez}, M.~R. 1994, \aaps, 104

\bibitem[{{Torres} {et~al.}(2018){Torres}, {Cidale}, {Kraus}, {Arias},
  {Barb{\'a}}, {Maravelias}, \& {Borges Fernandes}}]{2018A&A...612A.113T}
{Torres}, A.~F., {Cidale}, L.~S., {Kraus}, M., {et~al.} 2018, \aap, 612, A113,
  \dodoi{10.1051/0004-6361/201731723}

\bibitem[{{Vink} {et~al.}(2002){Vink}, {Drew}, {Harries}, \&
  {Oudmaijer}}]{2002MNRAS.337..356V}
{Vink}, J.~S., {Drew}, J.~E., {Harries}, T.~J., \& {Oudmaijer}, R.~D. 2002,
  \mnras, 337, 356, \dodoi{10.1046/j.1365-8711.2002.05920.x}

\bibitem[{{Vioque} {et~al.}(2018){Vioque}, {Oudmaijer}, {Baines},
  {Mendigut{\'\i}a}, \& {P{\'e}rez-Mart{\'\i}nez}}]{2018A&A...620A.128V}
{Vioque}, M., {Oudmaijer}, R.~D., {Baines}, D., {Mendigut{\'\i}a}, I., \&
  {P{\'e}rez-Mart{\'\i}nez}, R. 2018, \aap, 620, A128,
  \dodoi{10.1051/0004-6361/201832870}

\bibitem[{{Wheelwright} {et~al.}(2012){Wheelwright}, {de Wit}, {Weigelt},
  {Oudmaijer}, \& {Ilee}}]{2012A&A...543A..77W}
{Wheelwright}, H.~E., {de Wit}, W.~J., {Weigelt}, G., {Oudmaijer}, R.~D., \&
  {Ilee}, J.~D. 2012, \aap, 543, A77, \dodoi{10.1051/0004-6361/201219325}

\bibitem[{{Wouterloot} {et~al.}(1988){Wouterloot}, {Brand}, \&
  {Henkel}}]{1988A&A...191..323W}
{Wouterloot}, J.~G.~A., {Brand}, J., \& {Henkel}, C. 1988, \aap, 191, 323

\bibitem[{{Zickgraf}(2003)}]{2003A&A...408..257Z}
{Zickgraf}, F.-J. 2003, \aap, 408, 257, \dodoi{10.1051/0004-6361:20030999}

\bibitem[{{Zickgraf} {et~al.}(1986){Zickgraf}, {Wolf}, {Stahl}, {Leitherer}, \&
  {Appenzeller}}]{1986A&A...163..119Z}
{Zickgraf}, F.~J., {Wolf}, B., {Stahl}, O., {Leitherer}, C., \& {Appenzeller},
  I. 1986, \aap, 163, 119

\bibitem[{{Zickgraf} {et~al.}(1985){Zickgraf}, {Wolf}, {Stahl}, {Leitherer}, \&
  {Klare}}]{1985A&A...143..421Z}
{Zickgraf}, F.~J., {Wolf}, B., {Stahl}, O., {Leitherer}, C., \& {Klare}, G.
  1985, \aap, 143, 421

\end{thebibliography}

\end{document}